\documentclass[onecolumn,11pt]{article}
\usepackage[top=1in, bottom=1in, left=1in, right=1in]{geometry}
\usepackage{amsmath,amsfonts,amscd,amssymb}
\usepackage{graphicx}
\usepackage{epstopdf}
\usepackage{overpic}
\usepackage{cancel}
\usepackage{rotating}
\usepackage{url}
\usepackage{caption}
\usepackage{color}
\usepackage{rotating}
\usepackage{multirow}
\usepackage{wrapfig}
\usepackage{mathtools}
\usepackage{subeqnarray}
\usepackage{setspace}
\usepackage{palatino} 
\usepackage[numbers,sort&compress]{natbib}
\usepackage{framed}
\usepackage{xcolor}
\usepackage{enumitem}

\usepackage[bottom,flushmargin,hang,multiple]{footmisc}
\usepackage{lipsum}
\newcommand\blfootnote[1]{%
  \begingroup
  \renewcommand\thefootnote{}\footnote{#1}%
  \addtocounter{footnote}{-1}%
  \endgroup
}

\definecolor{header1}{cmyk}{0,0,0,1}

\title{\vspace{-.55in}{\fontsize{16}{16}\selectfont \textbf{Phase-based control of periodic fluid flows}}\vspace{-.15in}}

\author{\normalsize{Aditya G.~Nair$^{1*}$, Kunihiko Taira$^{2}$, Bingni W. Brunton$^3$, Steven L. Brunton$^1$}\\
\footnotesize{$^1$ Department of Mechanical Engineering, University of Washington, Seattle, WA 98195}\\
\footnotesize{$^2$ Department of Mechanical and Aerospace Engineering, University of California,
Los Angeles, CA 90095}\\
\footnotesize{$^3$ Department of Biology, University of Washington, Seattle, WA 98195\vspace{-.2in}}\\
}
\date{}
\begin{document}
\maketitle

\blfootnote{$^*$ Corresponding author (agnair@uw.edu).\\ 
\noindent \textbf{Code:}  https://github.com/nairaditya/Phase-control-fluid-flow}
\vspace{-.2in}

\begin{abstract}
Fluid flows play a central role in scientific and technological development, and many of these flows are characterized by a dominant oscillation, such as the vortex shedding in the wake of nearly all transportation vehicles.  
The ability to control vortex shedding is critical to improve the aerodynamic performance of these unsteady fluid flow systems.  
This goal requires precise characterization of how perturbations affect the long-time phase of the oscillatory flow, as well as the ability to control transient behaviors.  
In this work, we develop an energy-efficient flow control strategy to rapidly alter the oscillation phase of time-periodic fluid flows, leveraging theory developed for periodic biological systems. 
First, we perform a phase-sensitivity analysis to construct a reduced-order model for the response of the flow oscillation to impulsive control inputs at various phases. 
Next, we introduce two control strategies for real-time phase control based on the phase-sensitivity function: 1) optimal phase control, obtained by solving the Euler-Lagrange equations as a two-point boundary value problem, and 2) model-predictive control (MPC). 
Our approach is demonstrated for two unsteady flow systems, the incompressible laminar flow past a circular cylinder and the flow past an airfoil.  
We show that effective phase control may be achieved with several actuation strategies, including blowing and rotary control.  
Moreover, our control approach uses realistic measurements of the lift force on the body, rather than requiring high-dimensional measurements of the full flow field.  

\end{abstract}

\section{Introduction}

Periodic fluid flows are ubiquitous, for example in tidal flows, pulsatile flow of blood, 
rotating flows in turbomachines and engines, and vortex shedding in the wake of transportation vehicles.  
Controlling fluid flows~\cite{Bewley2001pas,Bewley:07,Brunton2015amr}, and periodic vortex shedding specifically~\cite{colonius2011control}, has remained a significant engineering challenge, with tremendous potential impact for industry and technology. 
Despite the challenges of flow control, researchers have long been inspired and encouraged by the exceptional performance of biological flyers and swimmers that take advantage of unsteady aerodynamics for phenomenal maneuverability in a range of flow conditions~\citep{Ellington:Nature96,dickinson1999wing,Shyy08, Dabiri2009arfm,wu2011fish, ramananarivo2011rather}.  
To achieve this performance, even in high-amplitude gusting conditions, the animal must sense the flow and adjust its wings or fins in real-time~\citep{shyy2016aerodynamics}.  
For instance, \citet{dickinson1999wing} suggested that regulating the phase of wing rotation effectively controls the aerodynamic forces during maneuvers. 
The wing motion is, in turn, regulated by modifying the activation phase of the associated muscles~\cite{dickinson2000animals}, which involves precise timing modulation~\cite{sponberg2012abdicating}. 
Inspired by these natural systems, there are ongoing efforts to develop simple yet robust control laws for engineering flows.  

Most engineering systems rely on steady or quasi-steady analysis, for which well-established controllers can be designed and implemented within comfortably predefined ranges of uncertainty and perturbations~\citep{Bewley:07}.  
However, linearizing the flow about a steady state is often inadequate to precisely control large-amplitude flow oscillations and unsteadiness~\citep{Bewley2001pas}.
At the moment, there are few studies that perform control of flows with large unsteady fluctuations, due to the challenges of high-dimensionality, strong nonlinearity, and time-delays~\citep{Brunton2015amr, Brunton2019arfm}.    
One of the main approaches for analyzing perturbation dynamics about time-varying periodic base flows is Floquet analysis~\cite{herbert1987floquet,barkley1996three}.  
The transient instabilities in these time-varying flows can also be described using optimally time-dependent modes~\cite{babaee2016minimization}, which can then be leveraged for flow control~\cite{blanchard2019stabilization}.

Phase-reduction analysis~\cite{nakao2016phase, kuramoto2019concept} provides a complementary approach to Floquet theory, enabling a description of the phase of a dynamical system comprised of interacting nonlinear oscillators.
Phase-reduction analysis has a rich history, especially in neuroscience~\cite{brown2004phase, schultheiss2011phase}, dating back to the works of~\citet{malkin1949lyapunov,kuramoto1984chemical} and~\citet{winfree2001geometry}. 
The analysis identifies a set of points in phase space that result in oscillations having the same phase, called \emph{isochrons}~\cite{guckenheimer1975isochrons}.  
Similarly, amplitude reduction identifies a set of points that converge to the same trajectory, called \emph{isostables}~\cite{wilson2016isostable}.
These isochrons and isostables are eigenfunctions of the infinite-dimensional Koopman operator~\cite{Koopman1931pnas,Mezic2005nd,mauroy2013isostables}. 
Techniques for amplitude-phase reduction can also be extended to analyze transient responses far from the stable limit-cycle dynamics~\cite{shirasaka2017phase} and to build network models~\cite{nair2018networked}.
Further, these amplitude-phase response models have been widely used for control in the context of neuronal dynamics and circadian rhythms~\cite{efimov2009controlling}.
One such effort develops a closed-loop approach to tune deep brain stimulation to suppress essential tremors~\cite{holt2016phasic}.
Our work leverages phase-based analysis and control to manipulate the transient behavior in unsteady fluid flows. 

Transient flow control is a challenging problem due to the fundamental difficulty in (i) precisely characterizing the response to actuation due to the nonlinear, high-dimensional nature of the underlying physics, and (ii) developing control strategies over a time scale commensurate with variations in the flow. 
Phase reduction analysis alleviates the former concern as it characterizes the effect of perturbations on the phase dynamics of oscillatory systems.
The analysis is especially suitable for high-dimensional systems as it can collapse the full-state dynamics to a phase-based representation in a single scalar variable.
The collective phase description of spatially extended systems, such as oscillatory convection, was first introduced in the works of~\citet{kawamura2013collective} and~\citet{kawamura2015phase}.          
More recently,~\citet{taira2018phase} used a phase-reduction approach to analyze synchronization of perturbation dynamics about time-varying base states for unsteady wake flows. 
Using this approach, the conditions for synchronization (lock-on) to external excitation were determined~\cite{Munday:PF13}.
Such a phase-reduced representation of the flow physics forms the basis of the analysis in this work. 
Instead of determining the lock-on conditions, we deduce the optimal external excitation to achieve transient control of periodic fluid flows.

In this work, we demonstrate the ability to perform rapid, energy-efficient control of highly unsteady periodic flows using phase-sensitivity analysis.  
In particular, we leverage recent developments in variational methods to determine phase-based optimal control inputs that alter the transient behavior (synaptic activation time) of spiking neurons~\cite{moehlis2006optimal, monga2019phase}, extending them to enable high-dimensional transient flow control. 
We demonstrate our approach on several relevant flow systems, including the illustrative Stuart-Landau oscillator model, and the canonical flow examples of laminar flow past a cylinder and flow over an airfoil.  
In each example, we demonstrate the robustness of the proposed approach, as well as the ability to incorporate physically realistic measurements and actuation strategies.  
Based on the phase-reduction analysis, we introduce two control optimization strategies, based on the Euler-Lagrange equations and model predictive control. %
This formulation enables the real-time control of the phase of vortex shedding, including prescribing the desired time of vortex pinch-off.  
The ability to modify the phase dynamics of time-varying vortical flows can  accelerate or decelerate the associated vortex shedding process~\cite{Pastoor:JFM08, joe2011feedback}, which in turn affects the unsteady aerodynamic forces. 
Thus, we have provided a useful framework for transient control in oscillatory fluid flows, enabling the manipulation of aerodynamic characteristics in a time frame that is faster than the characteristic time of the flow oscillation.  

\section{Formulation}
\label{sec:formulation}

Our overarching goal is to characterize and control the phase of oscillatory fluid systems, which may be described as a dynamical system of the form
\begin{equation}
\dot{\boldsymbol{q}} = \boldsymbol{N}(\boldsymbol{q}) + \boldsymbol{f},
\label{eq0}
\end{equation}
where $\boldsymbol{q}$ is the oscillatory variable and $\boldsymbol{f}$ is the external forcing or control input. 
The fluid flows in this paper are governed by the non-dimensional, incompressible, forced Navier-Stokes equation
\begin{equation}
\frac{\partial \boldsymbol{q}}{\partial t}  = \underbrace{-\boldsymbol{q} \cdot \nabla \boldsymbol{q} + \nabla p + \frac{1}{Re} \nabla^2 \boldsymbol{q}}_{\boldsymbol{N}(\boldsymbol{q}(\boldsymbol{x},t))} + \boldsymbol{f}(\boldsymbol{x},t) ,~~~\nabla \cdot \boldsymbol{q} = \boldsymbol{0},
\label{eq1}
\end{equation}
where $\boldsymbol{q} = \boldsymbol{q}(\boldsymbol{x},t)$ is the velocity field, defined over a spatial coordinate $\boldsymbol{x}$ and time $t$, $p$ is the pressure, $Re$ is the Reynolds number, and $\boldsymbol{f} = \boldsymbol{f}(\boldsymbol{x},t)$ is the external forcing (actuation).  
It is common to represent the flow field discretely over a high-resolution spatial domain, resulting in a high-dimensional flow state $\boldsymbol{q}$ whose evolution is described by a system of coupled nonlinear ordinary differential equations.  
In these high-dimensional fluid flow systems, the oscillatory variable and forcing both depend on space and time. 
For unforced flows, $\boldsymbol{f} = \boldsymbol{0}$. 

A large class of high-dimensional unsteady fluid flows have time-periodic behavior
\begin{equation}
\boldsymbol{q}(\boldsymbol{x},t) = \boldsymbol{q}(\boldsymbol{x},t+T),
\end{equation}
where $T$ is the period of the flow $\boldsymbol{q}$ and $\omega_n = 2\pi/T$ is the corresponding natural frequency of oscillation. We focus on such time-periodic unsteady fluid flows in this work. 

Many flow systems are oscillatory because of periodic excitation, such as when the regular flapping of a bird's wings establishes a periodically varying flow field.   
In this case, it is possible to write the forcing as $\boldsymbol{f}(\boldsymbol{x},t) = \boldsymbol{f}(\boldsymbol{x},\sin(\omega_n t)) + \boldsymbol{f}'(\boldsymbol{x},t)$, where $\boldsymbol{f}(\boldsymbol{x},\sin(\omega_n t))$ is the periodic forcing and $\boldsymbol{f}'$ is the actuation that we have control over.  
For notational simplicity, we will subsume any external periodic forcing $\boldsymbol{f}$ into the nonlinear dynamics in \eqref{eq0} and consider $\boldsymbol{f}$ to represent our actuation.  

\begin{figure}
  \centerline{\includegraphics[width=1.0\textwidth]{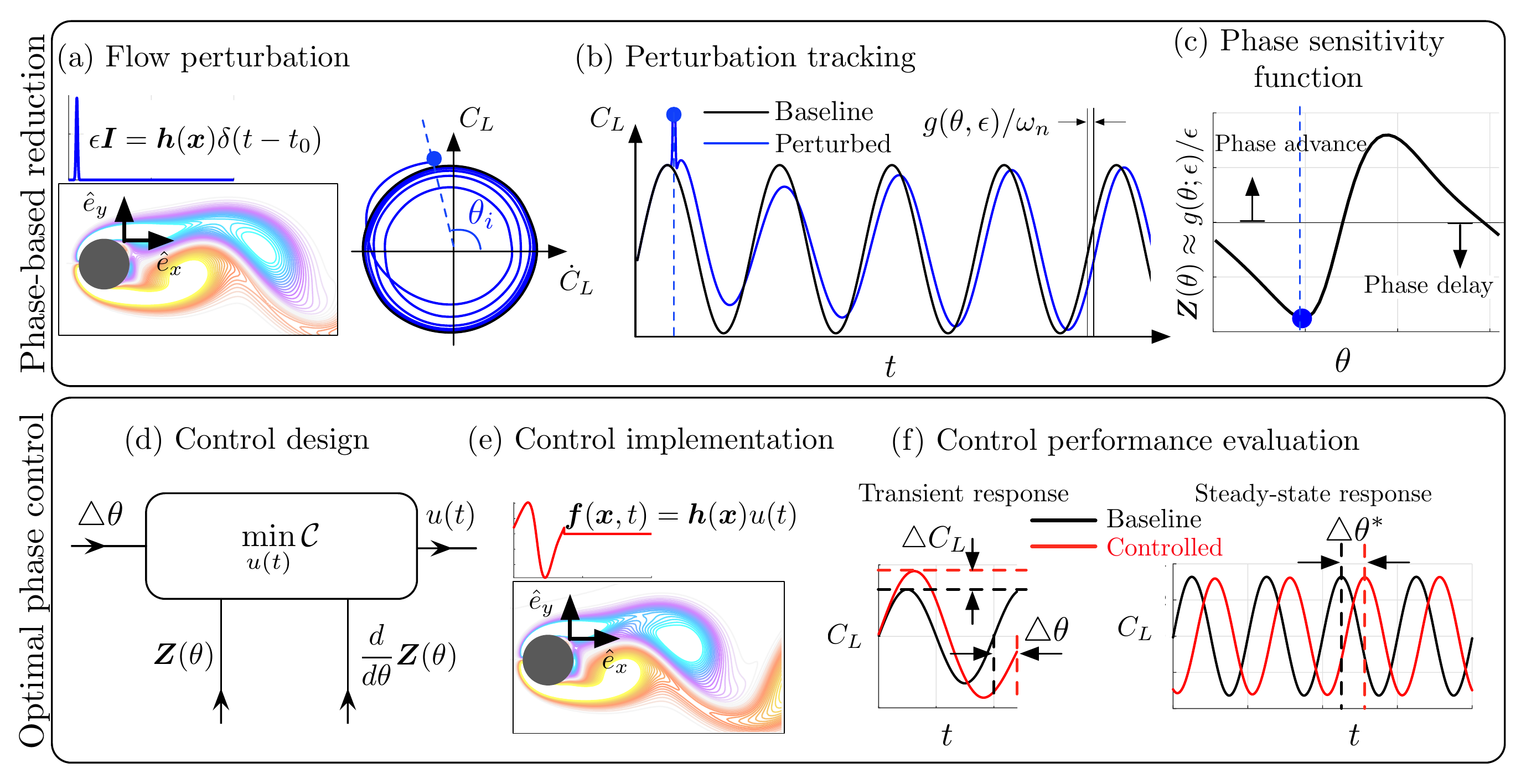}}
  \caption{Overview of phase reduction analysis: (a) Introduction of impulse flow perturbation, (b) baseline and perturbed lift coefficient trajectories (left: phase portrait of oscillations in $\dot{C}_L$ - $C_L$ plane, right: time history) and (c) phase sensitivity function $\boldsymbol{Z}(\theta)$. Overview of energy-optimal phase control: (d) Design of optimal control input $u(t)$ for desired phase shift $\triangle \theta$ and (e) controlled flow with the forcing $\boldsymbol{f}(t)$, (f) transient and steady-state response of lift coefficient on application of control.}
\label{fig1}
\end{figure}

This work designs an energy-optimal forcing $\boldsymbol{f}(\boldsymbol{x},t)$
to control the oscillatory phase of time-periodic fluid flows.  
Our work is inspired by two threads of prior work: phase-reduction analysis for high-dimensional fluid flows~\cite{taira2018phase} and energy-optimal phase control for neuronal systems~\cite{moehlis2006optimal}. 
Combining these approaches results in a strategy for energy-optimal phase control of unsteady fluid flows, summarized in Figure~\ref{fig1}. The overall procedure is carried out in two stages as follows:
\begin{enumerate}
\item \textbf{Phase-reduction analysis:} Following the work of~\citet{taira2018phase}, we extend phase-reduction analysis~\cite{nakao2016phase} to describe the high-dimensional behavior of periodic wake flows.  
The high-dimensional, unsteady fluid flow governed by Eq.~(\ref{eq1}) is reduced to a phase-based representation. 
This stage requires (a) introducing perturbations to the flow at various phases of oscillation, (b) tracking the effect of these perturbations compared to baseline oscillations, and (c) constructing a reduced-order model described by the phase-sensitivity function. 
The phase-based reduced-order model describes the phase dynamics of oscillations in the unsteady fluid flow system. 
This modeling procedure is explained in more detail in \S\ref{subsec:pra}.
\item \textbf{Optimal phase control:}
In this stage, we extend the optimal control strategy developed by~\citet{moehlis2006optimal} to high-dimensional systems, such as fluid flows. 
In particular, we alter the characteristic phases of unsteady fluid flow systems on a timescale that is commensurate with the characteristic flow timescale. 
This characteristic time corresponds to the natural shedding frequency in unsteady bluff-body flows. 
This stage includes (d) an optimal open-loop control design strategy using the Euler-Lagrange equations to determine the control input required for desired phase alteration of the flow, (e) implementation of the control law in the unsteady fluid flow, and (f) evaluation of the transient and steady-state control performance along with the associated flow physics.
The open-loop control forcing is designed offline using the phase-reduced model and is then introduced to the flow; however, it is also possible to use the reduced-order model for real-time feedback control with model predictive control. 
This optimal phase control procedure is described in detail in \S\ref{subsec:eopc}.
\end{enumerate}

Before we describe the formulation for controlling oscillatory phase in detail, we outline some of the goals and objectives of this work. In terms of the methodology, we extend the optimal phase-based control formulation of~\citet{moehlis2006optimal} to high-dimensional dynamical systems, in particular, for unsteady fluid flows. An alternative closed-loop model predictive control strategy is also proposed. Some of the specific physics-based objectives are to (i) achieve a transient phase shift in unsteady fluid flow oscillations by a desired amount after implementation of control,  (ii) explore the changes in the maximum lift coefficient in the immediate transient after application of control for varying phase alterations, (iii) examine the asymptotic phase shift achieved in steady-state, long after the control is switched off, and (iv) investigate the implications of phase control on the vortex formation and shedding process. 

\subsection{Phase-reduction analysis}
\label{subsec:pra}

We first apply phase-reduction analysis~\cite{kuramoto1975self, kuramoto2019concept} to a general class of time-periodic dynamical systems, given by Eq.~(\ref{eq0}). 
The phase of the time-periodic solution $\boldsymbol{q}$ of the unforced system, with frequency $\omega_n$, is given by
\begin{equation}
\dot{\theta}(t) = \omega_n,~~~ \theta \in [-\pi,\pi],
\end{equation}
where $\theta$ is the phase of the limit-cycle oscillation. 
The single scalar phase variable $\theta$ describes the oscillatory behavior of $\boldsymbol{q}$.
Solutions $\boldsymbol{q}$ near the limit cycle will all have frequency $\omega_n$, although they will have different phases, given by the scalar field $\Theta(\boldsymbol{q})$, with dynamics given by 
\begin{equation}
\dot{\Theta}(\boldsymbol{q}) = \nabla_{\boldsymbol{q}} \Theta(\boldsymbol{q}) \cdot \dot{\boldsymbol{q}} = \nabla_{\boldsymbol{q}} \Theta(\boldsymbol{q}) \cdot \boldsymbol{N}(\boldsymbol{q}) = \omega_n.
\end{equation}
The level sets of the phase function $\Theta(\boldsymbol{q})$ constitute the isochron coordinates, which are directly perpendicular to $\nabla_{\boldsymbol{q}} \Theta(\boldsymbol{q})$~\cite{guckenheimer1975isochrons}. 

To deduce the phase-reduction of the full system given by Eq.~(\ref{eq0}), we consider the introduction of small perturbations to the system at characteristic phases $\theta$ such that $\boldsymbol{f}(\mathbf{x},t) =  \epsilon \boldsymbol{I}(\mathbf{x},t)=\boldsymbol{h}(\boldsymbol{x})\delta(t-t_0)$, where $\epsilon \ll1$, $\boldsymbol{h}(\mathbf{x})$ is a spatial mode with $\|\boldsymbol{h}\|=1$, and $\delta$ is a unit-area delta function centered at the time $t_0\in[0,T)$ when $\boldsymbol{q}$ is at phase $\theta$. Ignoring the higher-order effects we obtain  
\begin{equation}
\dot{\Theta}(\boldsymbol{q}) = \nabla_{\boldsymbol{q}} \Theta(\boldsymbol{q}) \cdot \dot{\boldsymbol{q}} =  \nabla_{\boldsymbol{q}} \Theta(\boldsymbol{q}) \cdot [\boldsymbol{N}(\boldsymbol{q}) +  \boldsymbol{f}] = \omega_n + \underbrace{\epsilon\nabla_{\boldsymbol{q}} \Theta(\boldsymbol{q}) \cdot  \boldsymbol{h}}_{g(\theta,\epsilon)}\delta(t-t_0).
\label{eq2}
\end{equation}
Including higher-order effects is related to center-manifold reduction, as discussed in~\citet{kuramoto2019concept}.
We can measure the asymptotic phase shift along the limit cycle for each perturbed initial phase $\theta$ to compute the phase sensitivity function $\boldsymbol{Z}(\theta)$ as 
\begin{equation}
 \boldsymbol{Z}(\theta) = {\text{lim}}_{{\epsilon \rightarrow 0}} \frac{g(\theta,\epsilon)}{\epsilon} 
 = \nabla_{\boldsymbol{q}} \Theta(\boldsymbol{q})|_{\theta}\cdot \boldsymbol{h}.
\label{eq3}
\end{equation}
Here, $g(\theta,\epsilon)$ is the difference in the phase of baseline and perturbed trajectories at steady-state. 
Note that there are alternative approaches to compute the phase sensitivity function $\boldsymbol{Z}(\theta)$. 
One approach is to solve an associated adjoint problem to Eq.~(\ref{eq2}) with Malkin's normalization condition~\cite{brown2004phase}. 
Global isochrons can also be computed using the method of Fourier averages~\cite{mauroy2012use, mauroy2018global}.

Once the phase sensitivity function is determined, the phase dynamics of the dynamical system can be approximated linearly through the phase-reduced model
\begin{equation}
\dot{\theta}(t) = \omega_n + \boldsymbol{Z}(\theta) \cdot \boldsymbol{f} = \omega_n + \epsilon \boldsymbol{Z}(\theta) \delta(t-t_0).
\label{eq4}
\end{equation}

Now, we apply this phase-reduction analysis to high-dimensional dynamical systems governed by Eq.~(\ref{eq1})~\cite{taira2018phase}.
Using the phase-reduction analysis described above, we can simplify the high-dimensional dynamics associated with periodic flows by introducing a phase coordinate $\theta$ that parameterizes the periodic behavior of $\boldsymbol{q}(\boldsymbol{x},t)$. 
Thus, the high-dimensional dynamics may be mapped to a low-dimensional representation of a phase coordinate $\theta$ via a phase function $\Theta(\boldsymbol{q})$.
To evaluate $\boldsymbol{Z}(\theta)$ given an actuation strategy with spatial mode $\boldsymbol{h}(\boldsymbol{x})$, we introduce small impulsive perturbations to the flow of the form,
\begin{equation}
\boldsymbol{I}(\boldsymbol{x},t): = \boldsymbol{h}(\boldsymbol{x})\delta(t-t_0) = \boldsymbol{h}(\boldsymbol{x})\frac{1}{\sqrt{2\pi}\sigma} \exp{\left[-\frac{1}{2}\left(\frac{t-t_0}{\sigma}\right)^2\right]},
\label{eq2_1}
\end{equation}
where $\boldsymbol{h}(\boldsymbol{x})$ is the spatial profile of perturbation, and $t_0$ corresponds to the time at a particular phase of the flow $\theta (t_0)$ when the impulse is introduced. 
For all cases considered, $\sigma = 10\triangle t$, where $\triangle t$ is the numerical time step. 
The spatial profile depend on the type of actuation considered.

Tracking these added perturbations to the high-dimensional flow system, we obtain the infinitesimal phase sensitivity function $\boldsymbol{Z}(\theta)$ in Eq.~(\ref{eq3}), evaluated at each perturbed initial phase. 
The phase sensitivity function characterizes the effect of external perturbations on the phase of the periodic limit-cycle orbit. 
Instead of tracking the entire high-dimensional flow field ($\boldsymbol{q}$), time-series data from sensor measurements can be used. 
In the case of unsteady time-periodic bluff body flows, we utilize the lift coefficient ($C_L$) to provide an accurate representation of the phase characteristics of the flow~\cite{williams2009lift, loiseau2018sparse, taira2018phase}.
The lift and drag coefficients are defined as 
\begin{equation}
C_L = \frac{F_\text{lift}}{\frac{1}{2}\rho U_\infty^2l}, ~~~ C_D = \frac{F_\text{drag}}{\frac{1}{2}\rho U_\infty^2l}
\end{equation}
where $F_\text{lift}$ and $F_\text{drag}$ are the lift and drag forces on the body, $\rho$ is the density, $U_\infty$ is the free stream velocity and $l$ is the characteristic length scale. Once, the phase-sensitivity function is obtained, we can formulate the reduced-order model in Eq.~(\ref{eq4}), which describes the phase evolution of infinitesimal perturbations on the flow.

A summary of the phase reduction analysis is outlined in Figure~\ref{fig1} (top panel). 
First, impulse response simulations are performed by introducing perturbations to the flow through our actuators. 
We explore several actuation strategies, including blowing/suction, rotation and pitching of the body.  
For illustration, we show one such strategy using blowing/suction actuation for the cylinder flow in Figure \ref{fig1}(a). 
Once the actuation strategy is determined, the evolution of the lift coefficient is tracked in the $\dot{C}_L$ - $C_L$ plane, as in Figure \ref{fig1}(b). 
The phase coordinate $\theta$ can be easily defined in this plane. 
We show the baseline limit cycle trajectories (in black) as well as the perturbed trajectories in blue. 
The impulsive perturbation is introduced at $\theta_i$. 
We measure the asymptotic phase shift $g(\theta,\epsilon)$ by comparing the baseline and perturbed lift trajectories.
Repeating the impulse response analysis for every $\theta$, we can compute the phase sensitivity function using Eq. (\ref{eq3}), as seen in Figure \ref{fig1}(c). 
As this method only requires the perturbed trajectories at steady-state compared to the baseline, the phase sensitivity function can be obtained in experiments as well~\cite{williams2009lift}.  
For $\boldsymbol{Z}(\theta) > 0$, phase delay in the perturbed trajectories is observed, i.e., the phase of the perturbed trajectories lag the baseline flow trajectories.  
For  $\boldsymbol{Z}(\theta) < 0$, phase advancement is observed with the perturbed trajectories leading the baseline trajectories in phase.

\subsection{Energy-optimal phase control}
\label{subsec:eopc}

In the previous section, we derived a reduced-order model to describe the phase dynamics of high-dimensional unsteady fluid flows. 
In this section, we manipulate these phase dynamics to advance or delay the phase of flow oscillation. 
We emphasize that the present control strategy aims to rapidly change the phase of the flow oscillations within a time scale commensurate with the characteristic time of flow oscillation $T$, rather than asymptotic phase control. 

Our specific goal is to determine the optimal forcing input $\boldsymbol{f}(\boldsymbol{x},t)$ in Eq.~(\ref{eq1}) to shift the phase of oscillation by a desired amount $\triangle \theta$ within a single period of oscillation $T$. 
This objective may be formulated as modifying the period of the first oscillation to be $T^*= T+\triangle \theta/\omega_n$.
After this first period, we turn off the control, and the flow returns to its original period, but with a fixed phase difference $\triangle \theta$.  
The phase shift is given by $\triangle \theta = \omega_n(T^*-T)$, so a phase delay corresponds to $T^*<T$ and a phase advance corresponds to $T^*>T$. 

The control signal will use the same actuation strategy as the perturbations in  Eq.~(\ref{eq2_1}), with spatial forcing $\boldsymbol{h}$, but with a non-impulsive control magnitude $u(t)$:
Thus, the forcing input added to the unsteady fluid flow is given by 
\begin{equation}
\boldsymbol{f}(\boldsymbol{x},t) = \left\{ 
\begin{array}{ll}
\boldsymbol{h}(\boldsymbol{x})u(t) & \text{if~} 0 \le t < T^*\\
0   & \text{if~} t \geq T^*.
\end{array}
\right.
\label{eq9}
\end{equation}
Here, the control signal $u(t)$ can take arbitrary shape as opposed to the impulsive perturbation considered above. 
Moreover, we are interested in the control signal $u(t)$ that minimizes the overall energy input to the high-dimensional fluid flow system, while achieving the desired phase control. 
For a specified time $T^*$, and for all control inputs $u(t)$ which evolve  from $\theta(0) = 0$ to $\theta(T^*) = 2\pi$, we seek the control signal that minimizes a cost function~\cite{moehlis2006optimal} of the form,
\begin{equation}
\mathcal{C}[u(t)] = \int_{0}^{T^*} \left\{ [u(t)]^2 + \lambda(t)[\dot{\theta}(t) - \omega_n - \boldsymbol{Z}(\theta) u(t)]\right\} dt,
\label{eq5}
\end{equation}
where $\lambda(t)$ is the Lagrange multiplier that enforces the system dynamics $\dot{\theta}(t) = \omega_n + \boldsymbol{Z}(\theta) u(t)$.  
The phase delay $\triangle \theta$ is automatically enforced by the boundary conditions. 

To minimize the cost function $\mathcal{C}[u(t)]$, we apply the calculus of variations~\cite{moehlis2006optimal, monga2019optimal, monga2019phase} and solve the associated Euler-Lagrange equations given by,
\begin{equation}
u(t) = \frac{\lambda(t)\boldsymbol{Z}(\theta(t))}{2}, \quad
\dot{\theta} = \omega_n + \frac{\lambda(t)[\boldsymbol{Z}(\theta)]^2}{2}, \quad
\dot{\lambda} = -\frac{[\lambda(t)]^2\boldsymbol{Z}(\theta)}{2} \frac{d}{d \theta}\boldsymbol{Z}(\theta)
\label{eq8}
\end{equation}
subject to initial and final conditions for phase, i.e.,  $\theta(0) = 0$ and $\theta(T^*) = 2\pi$.
This constitutes a two-point boundary value problem in the phase coordinate $\theta(t)$. 
Given the phase sensitivity function $\boldsymbol{Z}(\theta)$ and its gradient $d\boldsymbol{Z}(\theta)/d \theta$, Eq.~(\ref{eq8}) can be solved to deduce the optimal control input $u(t)$ for a desired phase shift $\triangle \theta$. 
If $\lambda$ is approximately constant, the control strategy is directly proportional to the phase sensitivity function, which is related to opposition control
~\cite{siegel2003feedback,gerhard2003model}.

In addition to the initial condition for phase, we also need the initial condition for the Lagrange multiplier $\lambda(0)$ to solve Eq.~(\ref{eq8}). 
The appropriate initial condition can be found by a shooting method. 
Using the converged initial conditions, the optimal control signal $u(t)$ can be deduced. 
We can then compute the optimal forcing input given by Eq.~(\ref{eq9}) which is subsequently added to the Navier-Stokes equation in Eq.~(\ref{eq1}) to achieve the desired phase shift. 

A summary of this phase control strategy is shown in Figure \ref{fig1} (bottom panel). 
Here, for a desired phase shift $\triangle \theta$, the phase sensitivity function and its gradient are utilized to deduce the  optimal control input $u(t)$ minimizing the cost function $\mathcal{C}[\boldsymbol{u}(t)]$, as shown in Figure \ref{fig1}(d). 
The control forcing is then added to the Navier-Stokes equation as shown in Figure \ref{fig1}(e) after which the transient and steady-state control performance is evaluated, as in Figure \ref{fig1}(f).

An alternative to the phase control strategy described above is to use model predictive control (MPC)~\cite{garcia1989model, morari1999model} based on the phase response model.
Instead of solving the Euler-Lagrange equations, an optimal control problem can be solved in real-time over a receding horizon subject to the imposed system constraints. 
The procedure for MPC-based phase-optimal control is described in the appendix.
Using a simple example, we aim to show the connections between the feedback law derived using MPC and that of the energy-optimal phase control strategy described earlier in this section. 
The energy-optimal control strategy provides us with an intuition regarding the nature of the optimal control input by directly relating it to the phase sensitivity function. 
MPC does not provide such an intuitive connection, and obtaining an effective control law requires tuning of state and input weights. 
However, noise robustness and the ability to include constraints make MPC an attractive alternative strategy for phase control of fluid flows.  This robustness is especially for systems that have stochastic disturbances and slowly varying parameters, as MPC is able to correct for model uncertainty through sensor feedback.
Throughout this work, we primarily use the optimal control design based on the Euler-Lagrange equations, due to its simplicity and effectiveness. 
This approach is also compared to MPC  on a simple example in the next section, showing that performance is comparable.

\section{Results}
\begin{figure}
\vspace{-.1in}
  \centerline{\includegraphics[width=0.8\textwidth]{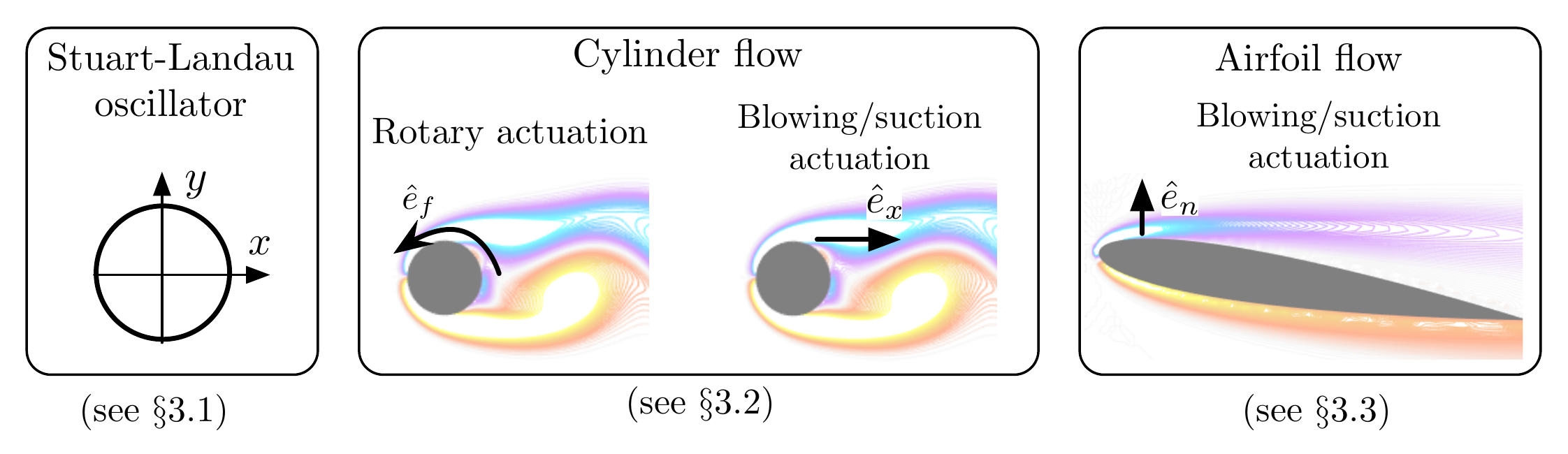}}
  \vspace{-.1in}
  \caption{Summary of the examples considered in this work for optimal phase control: (a) Stuart-Landau oscillator, (b) unsteady flow over a cylinder at $Re = 100$ with rotary and blowing/suction ($x$-direction) actuation, (c) unsteady two-dimensional flow over an airfoil at $Re = 1000$ and angle of attack of $\alpha = 9^\circ$ with blowing/suction actuation.}
\label{fig2}
\vspace{-.1in}
\end{figure}

We demonstrate the phase-based flow control strategy on three example systems as summarized in Figure \ref{fig2}. 
First, we investigate a Stuart-Landau oscillator, where the phase sensitivity function can be computed analytically. 
We then demonstrate the phase reduction analysis and optimal control on a canonical flow over a cylinder. 
Here, we explore two actuation strategies, one using rotary actuation and the other using momentum injection. 
Finally, we implement optimal phase control on a two-dimensional flow past an airfoil using blowing/suction that is normal to the flow. 

\subsection{Stuart-Landau equation}
\label{subsec:SL}

We begin with a simple motivating example, given by the Stuart-Landau (SL) oscillator
\begin{equation}
\frac{\mathrm{d}A}{\mathrm{d}t} = 
{a_0 A - a_1 A |A|^2}+  \sigma\xi(t)   + \text{Re}[f(t)],
\end{equation}
where $A(t) = x(t)+iy(t)$ is the complex amplitude, $f(t)$ is the forcing, and ${\xi(t) = \xi_x(t) + i\xi_y(t)}$ is independent and identically distributed Gaussian white noise with unit variance and strength $\sigma$. 
For $\sigma>0$, the SL equation exhibits stochastic dynamics.
We use a stochastic Runga-Kutta solver to integrate the dynamics~\cite{kasdin1995discrete}. 

\begin{figure}
  \centerline{\includegraphics[width=0.9\textwidth]{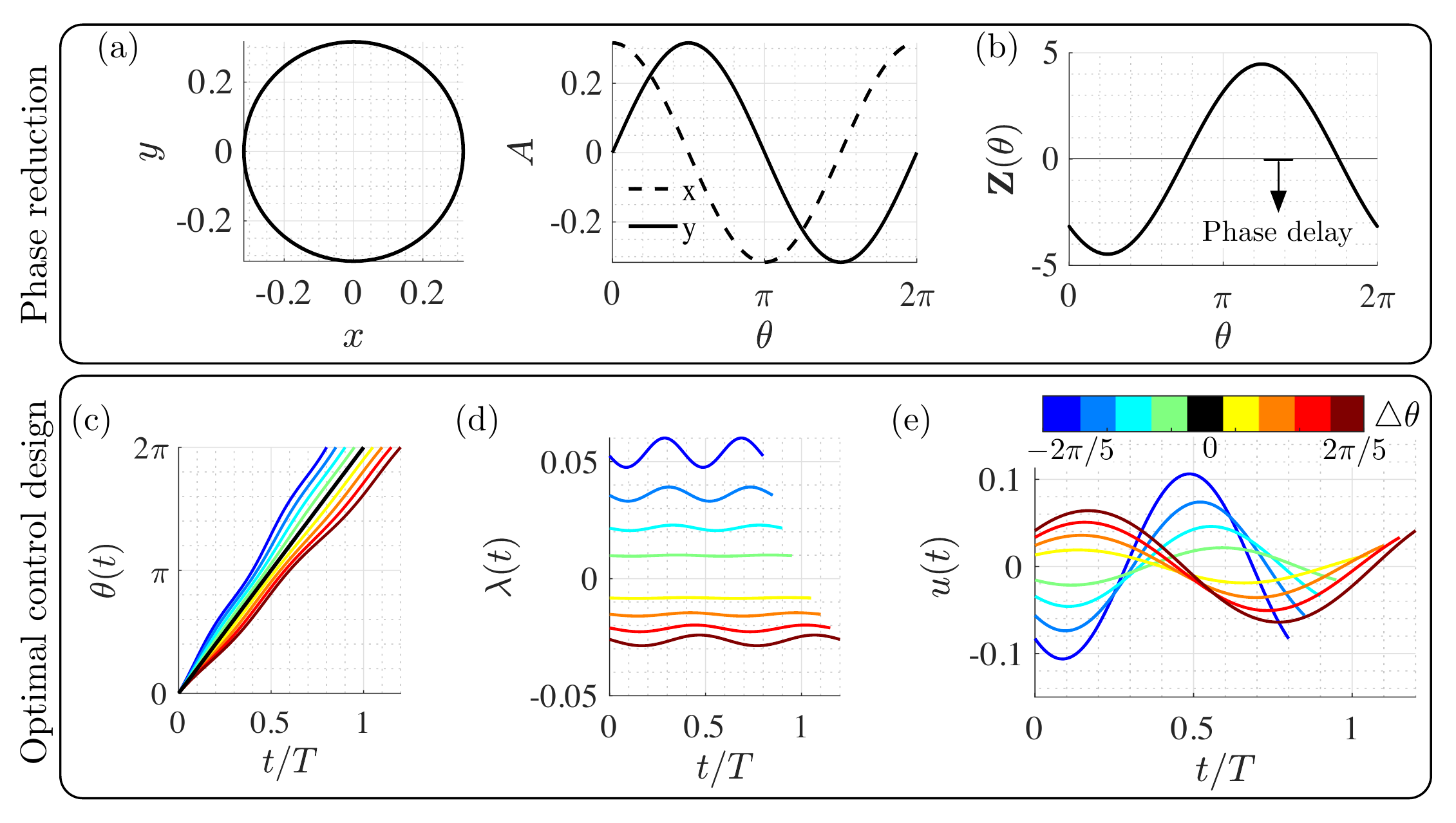}}
\vspace{-.1in}
\caption{Overview of phase-reduction and optimal phase control for a Stuart-Landau oscillator: (a) Baseline trajectories; (b) analytical phase-sensitivity function and optimal control design for desired phase shifts of $-2\pi/5 \le \triangle \theta \le 2\pi/5$; (c) phase coordinate dynamics; (d) dynamics of the Lagrange multiplier; and (e) the optimal control input from the Euler-Lagrange equations.}
\label{fig3}
\end{figure}

The Stuart-Landau equation was initially used to study the nonlinear  instability of plane Poiseuille flow~\cite{stuart1960non} and was later extended for weakly nonlinear stability analysis of cylinder wakes and open cavity flows~\cite{sipp2007global}. 
\citet{bagheri2013koopman} showed that the leading Koopman eigenfunctions of the cylinder flow can be constructed using these coordinates. 
In Cartesian coordinates, we obtain the well-known Hopf normal form, given by
\begin{subequations}\label{eq:SLeq}
\begin{align}
\frac{dx}{dt} &= \mu x - \omega y - (x^2+y^2)(x - \beta y) + \sigma \xi_x(t) + f(t), \label{xeq}\\ 
\frac{dy}{dt} &= \omega y + \mu x - (x^2+y^2)(\beta x + y) + \sigma \xi_y(t),\label{yeq}
\end{align}
\end{subequations}
where, $a_0 = \mu + i \omega$ and $a_1 = 1+i\beta$. 
Here, $f(t)$ is the forcing added in the $x$-direction (for phase control). 
We set $\mu = 0.1, \omega = 1$ and $\beta = 1$. 
The time step for numerical integration is $\triangle t = 0.01$.
The radial amplitude is given by $r= \sqrt{x^2 + y^2}$.
For the case with no noise, the radial amplitude is constant, $r(t) = r_0$, with a period of oscillation $T = 6.98$. 
The phase coordinate on the limit-cycle is $\theta = \tan^{-1}(y/x)$. 
The phase sensitivity function for forcing in the $x$-direction can be constructed analytically~\cite{moehlis2014improving} as 
\begin{equation}
Z(\theta) = \frac{\beta}{\sqrt{\mu}}\cos\theta -\frac{1}{\sqrt{\mu}}\sin\theta.
\label{eqZ}
\end{equation}
This phase-sensitivity function only depends on the system parameters and the phase coordinate. 

The baseline trajectories and the phase sensitivity function for this example are shown in Figure \ref{fig3}. 
A maximum phase delay is observed around $\theta = \pi/5$, while a maximum phase advance is observed around $\theta = 6\pi/5$. 
We can also deduce its gradient analytically from Eq.~(\ref{eqZ}). 
With this information, we are able to design an optimal phase controller,  as described in \S \ref{subsec:eopc}. 

We consider a range of phase shifts so that $-2\pi/5 \le \triangle \theta \le 2\pi/5$.
A phase advance of ${\triangle \theta = -2\pi/5}$ corresponds to $T^*/T = 0.8$ and a phase delay of $\triangle \theta = 2\pi/5$ corresponds to $T^*/T = 1.2$. 
The phase coordinate $\theta(t)$, corresponding Lagrange multiplier $\lambda(t)$, and the optimal control input $u(t)$ obtained by solving the associated Euler-Lagrange equations in Eq.~(\ref{eq8}) are shown in Figure \ref{fig3}(c), (d) and (e), respectively.
The color of the lines correspond to the varying values of desired phase shift.
 As described before, the initial condition $\lambda(0)$ is obtained by performing a shooting method. 
 The converged initial conditions fall in the range $-0.026 \le \lambda(0) \le 0.0525$.  
 For phase advance (negative phase shifts), $\lambda(t) > 0$, while for phase delays (positive phase shift), $\lambda(t) < 0$.
 The baseline phase is shown by the solid black line. 

 \begin{figure}
 \vspace{-.1in}
  \centerline{\includegraphics[width=1.0\textwidth]{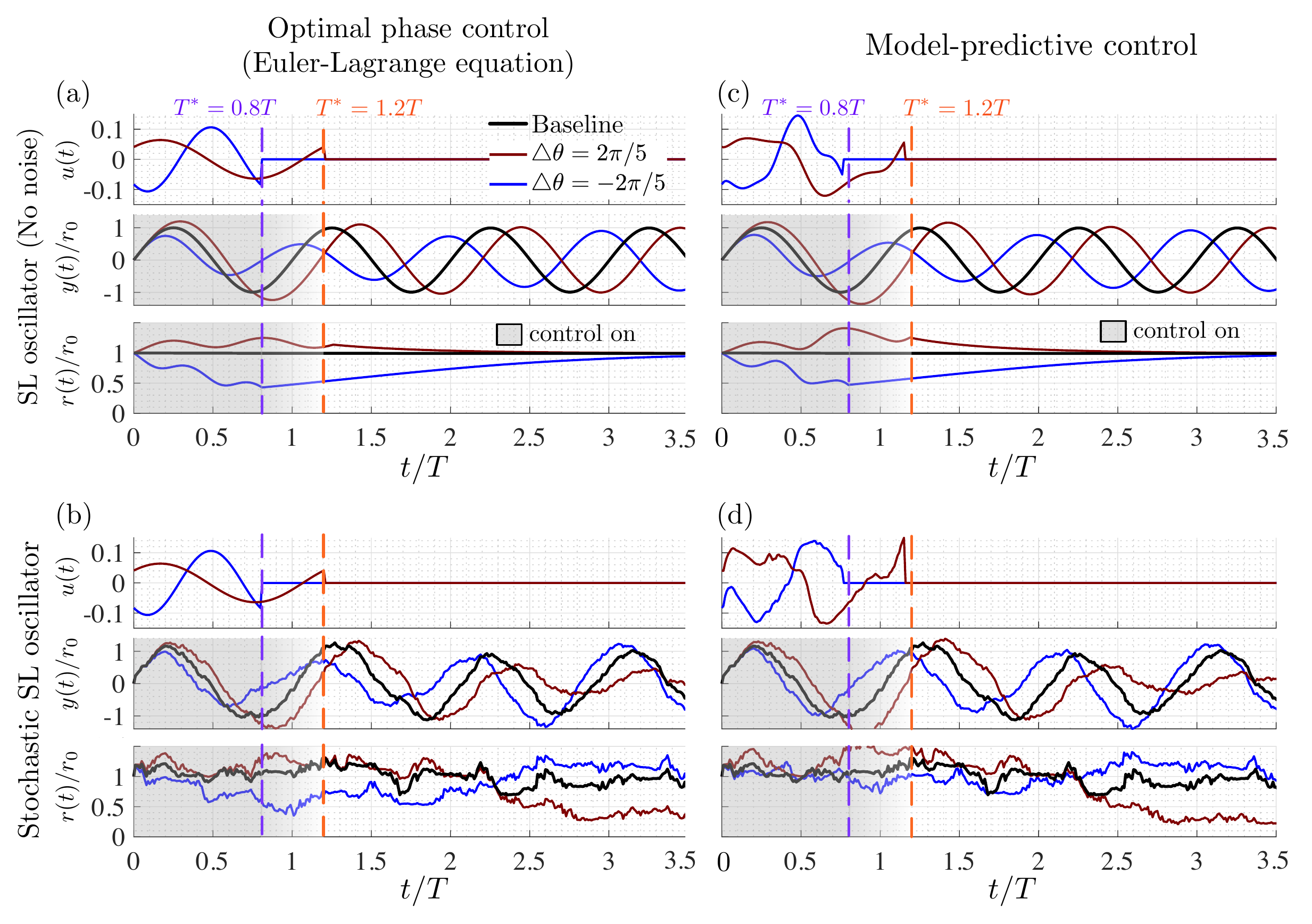}}
  \vspace{-.1in}
  \caption{Optimal phase control (using Euler Lagrange equation) of Stuart-Landau oscillator with (a) no noise ($\sigma = 0$) and (b) stochastic dynamics $\sigma = 0.05$. Model-predictive control of Stuart-Landau oscillator with (c) no noise ($\sigma = 0$) and (d) stochastic dynamics $\sigma = 0.05$. For each case, the control input $u(t)$, trajectory $y(t)$ and radial amplitude $r(t)$ are shown.}
\label{fig4}
\vspace{-.1in}
\end{figure}

We also notice the correspondence between the phase sensitivity function and the optimal control inputs $u(t)$ in Figure \ref{fig3}(b) and Figure \ref{fig3}(e), respectively. 
The shape of the control input closely follows that of the phase sensitivity function, especially for phase advance (negative phase shifts). 
For phase delay (positive phase shift), the control input follows the opposite trend of the phase sensitivity function.  
This is intuitive as the control input opposes any phase advances in the model by adding an opposite control force.   

We implement the optimal control law $f(t) = u(t)$ in the SL system in Eq.~(\ref{eq:SLeq}).
Control simulations are performed with no noise ($\sigma = 0$) and with stochastic dynamics ($\sigma = 0.05$) for extreme phase shifts of $\triangle \theta = \pm 2\pi/5$. 
The control input $u(t)$, control trajectories $y(t)$, and the radial amplitude $r(t) = \sqrt{x^2+y^2}$ for the baseline SL oscillator and the stochastic SL oscillator are shown in Figures \ref{fig4}(a) and (b), respectively. 
For the baseline SL oscillator, the prescribed phase shift is achieved in the immediate transient (within $T^*$) for both the control cases corresponding to $\triangle \theta = -2\pi/5$ and $\triangle \theta = 2\pi/5$.
This phase shift is also observed in the asymptotic behavior.
For the stochastic system, shown in Figure \ref{fig4}(b), the prescribed phase shifts are consistently achieved with control.
This demonstrates the robustness of the control technique to stochastic disturbances. 
However, the phase shifts are not maintained asymptotically after the control is switched off. 
When the control is on, along with the phase shift, we also notice an amplitude modulation in the trajectories of $y(t)$. 
For phase delay, an increase in the amplitude is observed while for phase advance, a decrease in the amplitude is observed. 
These changes in the amplitude by performing optimal phase control are clearly visible in $r(t)$ as well. 

Additionally, we perform real-time phase control using MPC, as discussed in the Appendix. 
The prediction and control horizons are fixed at $n_c = n_p = 8$, and we also fix $Q = 5, R = 1, R_{\triangle u} = 100$. 
The results using MPC are shown in Figure~\ref{fig4}(c) and (d). 
The MPC and energy-optimal results are remarkably similar. 
The variation in the control inputs follow similar trends for both control cases and responses are almost identical. 
The phase optimal control strategy not only helps to advance and delay phase, but it also appropriately modulates the amplitude characteristics. 
In the following examples, we will see the implications of optimal phase control applied to unsteady fluid flows.

\subsection{Laminar flow over a cylinder}
\label{subsec:CF}

We now demonstrate our phase control approach on the canonical two-dimensional, incompressible flow past a circular cylinder~\cite{Noack2003jfm} at Reynolds number of $Re = U_\infty d/\nu = 100$, where $U_\infty$ is the free stream velocity, $d$ is the cylinder diameter and $\nu$ is the kinematic viscosity.
Direct numerical simulations (DNS) are performed using the immersed boundary projection method~\citep{Colonius:CMAME08, kajishima2017numerical} with a computational grid extending over $(x/d,y/d) \in [-31,33] \times [-32, 32]$.
A multi-domain technique is utilized with smallest spatial resolution of $\triangle x/d = \triangle y/d = 0.02$. The  simulation  time step is $\triangle t = 0.005$. 
Details of the simulation and validation can be found in~\citet{Taira:JCP07}. 

The baseline flow exhibits periodic vortex shedding in the cylinder wake with shedding frequency of $\omega_n = 2\pi f_n U_\infty/d = 1.03$ (characteristic time $T = 2\pi/\omega_n = 6.1$). 
In this flow, we will explore two different actuation strategies, first using rotary actuation (\S~\ref{sec:rotary}) and then using momentum injection~(\S~\ref{sec:momentum}).
Before we introduce perturbations for phase-reduction analysis or forcing input for phase-based control, we need to develop the actuation strategies. 

\subsubsection{Rotary actuation}\label{sec:rotary}

In this section, we achieve phase control of the unsteady cylinder flow by rotating the cylinder as our actuation.  
Unlike the previous example of the SL oscillator, the phase-sensitivity function for this high-dimensional fluid flow cannot be deduced analytically. 
We follow the phase-reduction approach outlined in \S \ref{subsec:pra} to construct the phase sensitivity function. 
To introduce perturbations, we impulsively rotate the cylinder in the counter-clockwise direction, as in Eq. (\ref{eq2_1}).

\begin{figure}
\vspace{-.1in}
  \centerline{\includegraphics[width=0.95\textwidth]{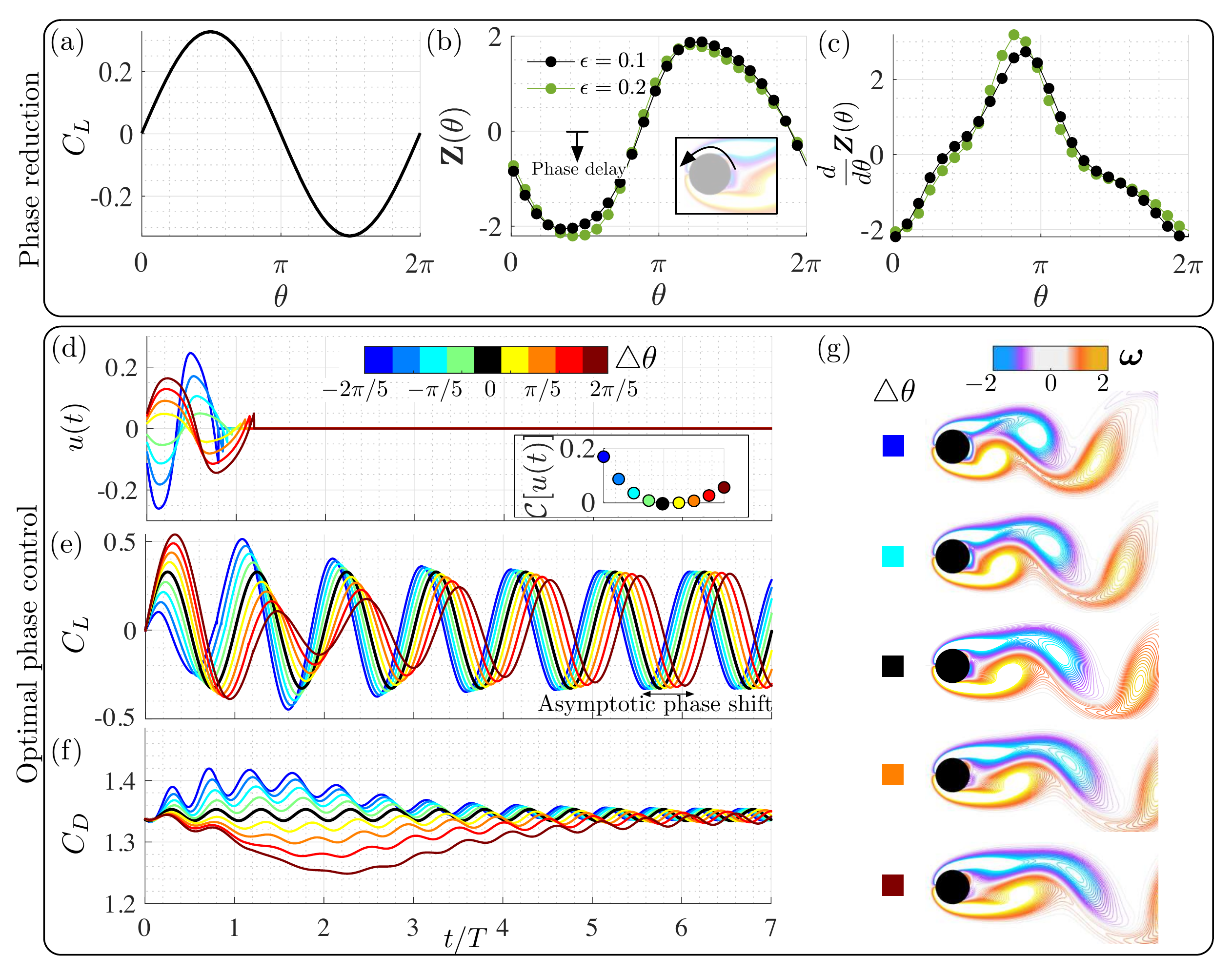}}
\vspace{-.1in}
\caption{Overview of phase-reduction and optimal phase control for cylinder flow using \textbf{rotary actuation}: (a) Lift coefficient ($C_L$) of baseline cylinder flow with respect to the phase coordinate $\theta$,  (b) phase sensitivity function and (c) its gradient (right) for impulse perturbation introduced at an actuator located near the top separation point in the $\hat{e}_x$ direction, (d) energy-optimal control input needed to shift the phase of oscillation (with associated cost function in inset), (e) lift and (f) drag characteristics after application of forcing in direct numerical simulations, and (g) instantaneous flow fields immediately after control is turned off ($t = T^*$). The color of the lines correspond to the cases with desired phase shift $\triangle \theta$ with $\triangle \theta = 0$ indicates baseline.}
\label{fig5}
\vspace{-.1in}
\end{figure}

For reference, we show the baseline lift fluctuation over a period of oscillation in Figure \ref{fig5}(a), which resembles the baseline trajectory of the Stuart-Landau oscillator. 
Multiple direct numerical simulations are performed by introducing weak rotary perturbations ($\epsilon = 0.1$) at various phases $\theta$. 
We then compute the asymptotic phase difference between the perturbed lift with respect to the baseline after $15$ periods of oscillation, resulting in the phase sensitivity function $\boldsymbol{Z}(\theta)$ using Eq. (\ref{eq3}).
We show the phase sensitivity function for perturbation size of $\epsilon = 0.1$ in Figure \ref{fig5}(b). 
For verification, we perform a similar analysis with a perturbation  size of $\epsilon = 0.2$, and both sensitivity functions are nearly identical. 
Generally, phase delay is observed for phases corresponding to $C_L >0$ and phase advance is observed for $C_L < 0$. 
The gradient of phase sensitivity ${d\boldsymbol{Z}(\theta)}/{d\theta}$ is shown in Figure \ref{fig5}(c). 
Larger gradients are observed at zero crossings of the lift curve.

Using the phase sensitivity function and its gradient, we are equipped to compute the optimal control input $u(t)$ for phase alteration of unsteady flow oscillations. 
We consider a range of desired phase shifts $-2\pi/5 \le \triangle \theta \le 2\pi/5$, which correspond to $0.8 T \le T^* \le 1.2 T$. 
Solving the Euler-Lagrange equations in Eq. (\ref{eq8}), the control inputs for this range of phase shifts are shown in Figure \ref{fig5}(d). 
As in the previous example, for phase advance (negative phase shift) the control input follows a similar trend to that of the phase sensitivity function, while for phase delay (positive phase shift) the control input follows the opposite trend.  
For positive phase shift, the control input opposes any phase advance in the sensitivity function.
The inset in Figure \ref{fig5}(d) shows the cost function $\mathcal{C}[u(t)]$ from Eq. (\ref{eq5}) corresponding to the different phase shifts. 
As expected, lower cost is associated with smaller phase shifts.
The cost for negative phase shifts is higher than for positive phase shifts, as the magnitude of input is larger. 

With the control input, we can determine the corresponding forcing to add to the Navier-Stokes equations, given by Eq. (\ref{eq9}). 
We perform direct numerical simulations with the optimal forcing for each case, and the control forcing is switched off after $t = T^*$.
The lift coefficients for the controlled simulations are shown in Figure \ref{fig5}(e). 
The baseline lift coefficient is shown in black.
For all control cases, immediately after the application of control, the desired phase shift is achieved. 
Thus, the flow immediately locks on to the frequency prescribed within characteristic time (one period).
In previous studies, the short-time Fourier transform and a Kalman filter were used to phase-lock the lift fluctuations with the control signal within 2-3 periods of oscillation~\cite{joe2011feedback}, while our controller achieves immediate phase-locking. 
The desired phase shift is sustained asymptotically after the control is switched off.

An important consequence of the phase-optimal control strategy is the effect on the amplitude of lift fluctuations in the immediate transient. 
We see that for positive phase shifts, the amplitude of lift fluctuation increases immediately after application of control.  
For the negative phase shift, in the first oscillation cycle, the lift amplitude decreases with control. 
However, we observe a sharp rise in the lift fluctuation over the baseline in the second oscillation cycle. 
This increase in amplitude was not observed in the Stuart-Landau oscillator. 
Along with the lift coefficient, we also show the drag coefficient in Figure~\ref{fig5}(f). 
The drag coefficient decreases with positive phase shifts while it increases with negative phase shifts in the immediate transient.
For positive phase shifts, a maximum drag reduction is achieved after the control is switched off. 
This is evidence that transient phase control is an effective means for both increasing lift and decreasing drag~\cite{amitay2002controlled}.
Though not shown, the transient phase control may be repeated after the system reaches steady state, about $6$ oscillation cycles later.

To understand the flow physics associated with the transient phase control, we analyze the instantaneous vorticity fields $\boldsymbol{\omega}(\boldsymbol{x},t)$ at $t= T^*$ for the different phase shifts, shown in Figure \ref{fig5}(g). 
We can see that for negative phase shifts, the vortex shedding takes place earlier and the length of the shed vortex is shorter. 
The size of the shed vortex is proportional to the magnitude of the desired phase shift.
Only once the vortex is shed, we see an increase in the lift fluctuation in the control cases of negative phase shifts (i.e. after $t = T^*$). 
Thus, one mechanism of enhancing lift is to shed the vortex earlier. 

The other mechanism of increasing lift is to delay the vortex shedding as much as possible, which is seen in the positive phase shift control cases. 
In these cases, the vortex shedding is delayed, resulting in an elongated vortical structure
This stabilizes the recirculation region behind the cylinder and leads to a drag reduction as well. 
Note that the cost associated with shedding a vortex earlier is slightly larger. 
We argue that for positive phase shifts, the control principle is similar to that of direct opposition control of vortices in the near wake, as discussed in the work of~\citet{siegel2003feedback} and~\citet{gerhard2003model}.
In these works, direct opposition control of wake structures with transversal oscillations of the cylinder enable the formation of elongated recirculation bubbles~\cite{Pastoor:JFM08}. 
Using a control law that opposes any phase advances in the phase-sensitivity function leads directly to control of vortices in the near wake.

\subsubsection{Momentum injection}\label{sec:momentum}

We now investigate the phase-optimal control technique for another actuation strategy based on momentum injection, where we add a localized force to the Navier-Stokes equation to mimic a blowing/suction type actuator. 
The spatial profile of the actuator $\boldsymbol{h}(\boldsymbol{x}) = \delta(\boldsymbol{x}-\boldsymbol{x}_s)$ is  a Dirac delta in space, approximated using the three-cell discrete delta function~\cite{Roma:JCP99}. 
The actuation is introduced at the average separation point $\boldsymbol{x}_s$ on the upper surface of the cylinder,  at an angle of $58.4^\circ$ from the aft separation point and positioned at $3\triangle x$ from the cylinder surface. 
The actuation is prescribed in the $x$-direction. 
While there is flexibility in choosing the actuation for the proposed control method, some actuation configurations may be more effective for controlling phase~\cite{taira2018phase, bhattacharjee2018optimal}.

\begin{figure}
\vspace{-.15in}
  \centerline{\includegraphics[width=0.95\textwidth]{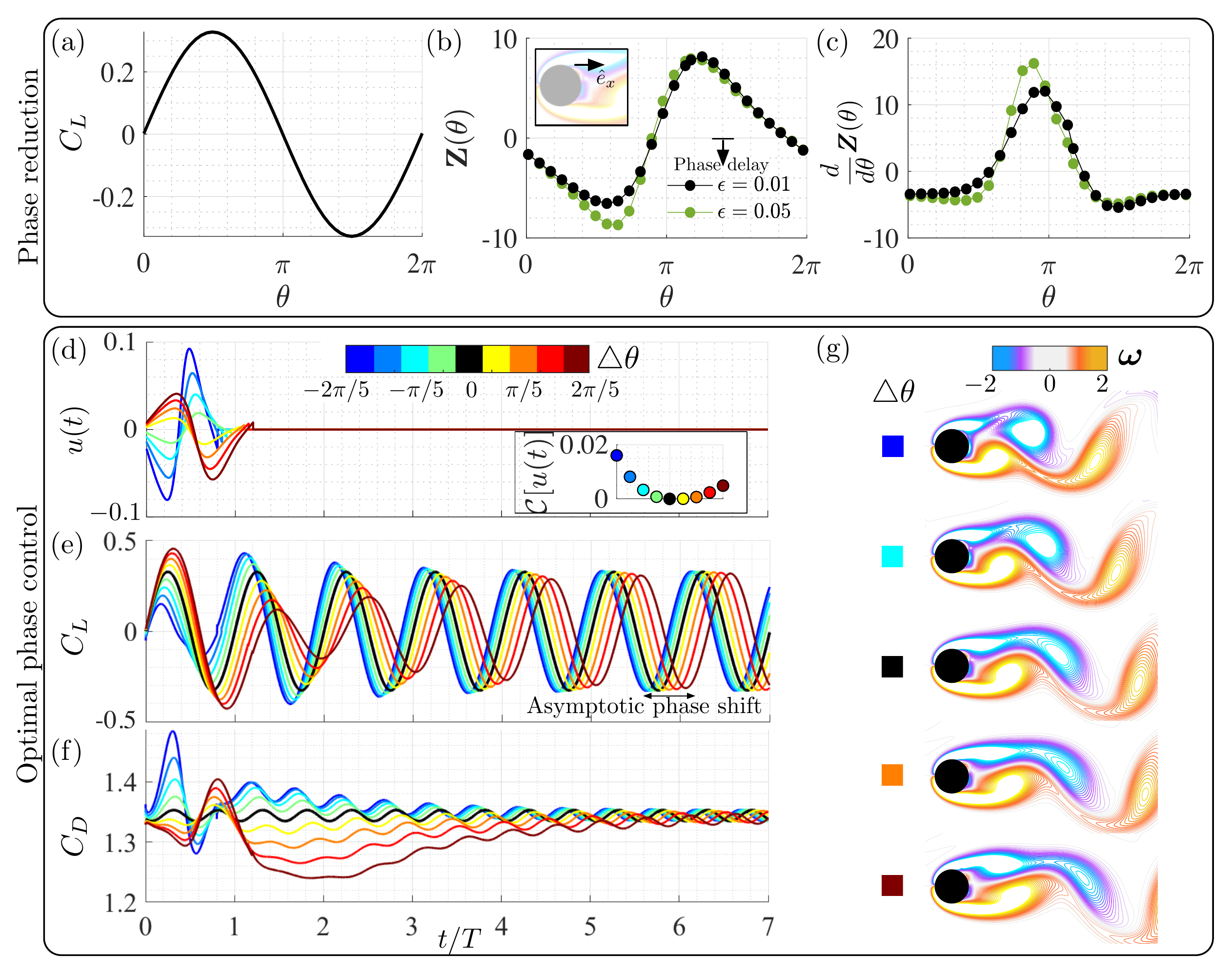}}
\vspace{-.15in}
\caption{Overview of phase-reduction and optimal phase control for cylinder flow using \textbf{momentum injection}: (a) Lift coefficient ($C_L$) of baseline cylinder flow with respect to the phase coordinate $\theta$,  (b) phase sensitivity function and (c) its gradient (right) for impulse perturbation introduced at an actuator located near the top separation point in the $\hat{e}_x$ direction, (d) energy-optimal control input needed to shift the phase of oscillation (with associated cost function in inset), (e) lift and (f) drag characteristics after application of forcing in direct numerical simulations, and (g) instantaneous flow fields immediately after control is turned off ($t = T^*$). The color of the lines correspond to the cases with desired phase shift $\triangle \theta$ with $\triangle \theta = 0$ indicates baseline.}
\label{fig6}\vspace{-.125in}
\end{figure}

For reference, the baseline lift is shown in Figure~\ref{fig6}(a).
Again, the phase sensitivity function $\boldsymbol{Z}(\theta)$ is obtained from impulsive perturbations, as shown in Figure \ref{fig6}(b). 
The analysis is performed for perturbation amplitudes of $\epsilon = 0.01$ and $0.05$. 
The phase sensitivity functions are similar for different perturbation sizes, except near $\theta = 3\pi/5$. 
Compared to the rotary control case, the phase sensitivity function has a larger magnitude at most characteristic phases.
This shows that the flow is more sensitive to momentum injection than rotary control. 
The gradient of the phase sensitivity function is highest around $\theta = \pi$, as seen in Figure~\ref{fig6}(c).

The phase control inputs to achieve the desired phase shifts using momentum injection are shown in Figure \ref{fig6}(d). 
As with rotary actuation, the cost function is larger for negative phase shifts (shown in inset). 
However, the cost of momentum injection is an order of magnitude lower. 
This is expected because of the increased phase sensitivity in the momentum injection case. 
We can also compute the actuation cost with the steady and unsteady momentum coefficients given by 
\begin{equation}
\bar{C}_\mu = \frac{\rho u_\text{jet}^2 \sigma}{\frac{1}{2}\rho U_\infty^2 d},~~~~~ {C}^\prime_\mu = \frac{\rho u_\text{rms}^2 \sigma}{\frac{1}{2}\rho U_\infty^2 d}, 
\label{eqcmu}
\end{equation}
respectively, where $\sigma$ is the actuator width, given by the grid spacing $\sigma = \triangle x$.
The characteristic velocities $u_\text{jet}$ and $u_\text{rms}$ are computed with companion simulations by setting the freestream velocity to be zero~\citep{Munday:PF13}.
The steady and unsteady coefficient of momentum for $\triangle \theta = -2\pi/5$ are $\bar{C}_\mu = 0.0078$ and $C^\prime_\mu = 0.0125$, respectively. For $\triangle \theta = 2\pi/5$, $\bar{C}_\mu = 0.006$ and $C^\prime_\mu = 0.01$.

Incorporating the forcing input in the Navier--Stokes equations, we perform direct numerical simulations for the varying desired phase shifts. 
The time history of the lift and drag coefficients in the controlled simulations are shown in Figure \ref{fig6}(e) and (f), respectively.
We see similar control performance with respect to the transient phase shift, lift fluctuation, and asymptotic phase shift compared to the rotary actuation case.
However, the performance in drag differs slightly. 
When the control in on, the drag both increases and decreases with respect to the baseline. 
When the control is off, the drag coefficient decreases for positive phase shift and increases slightly for negative phase shifts. 
This is consistent with the results from rotary actuation.  
The (controlled) instantaneous vorticity fields at $t = T^*$ are shown in Figure~\ref{fig6}(g).
The characteristic flow physics for the control based on momentum injection is almost identical to that of rotary actuation.
This shows the consistency of the optimal phase control approach in terms of altering the vortex shedding behavior for different actuation strategies.

\begin{figure}
\vspace{-.1in}
  \centerline{\includegraphics[width=1.0\textwidth]{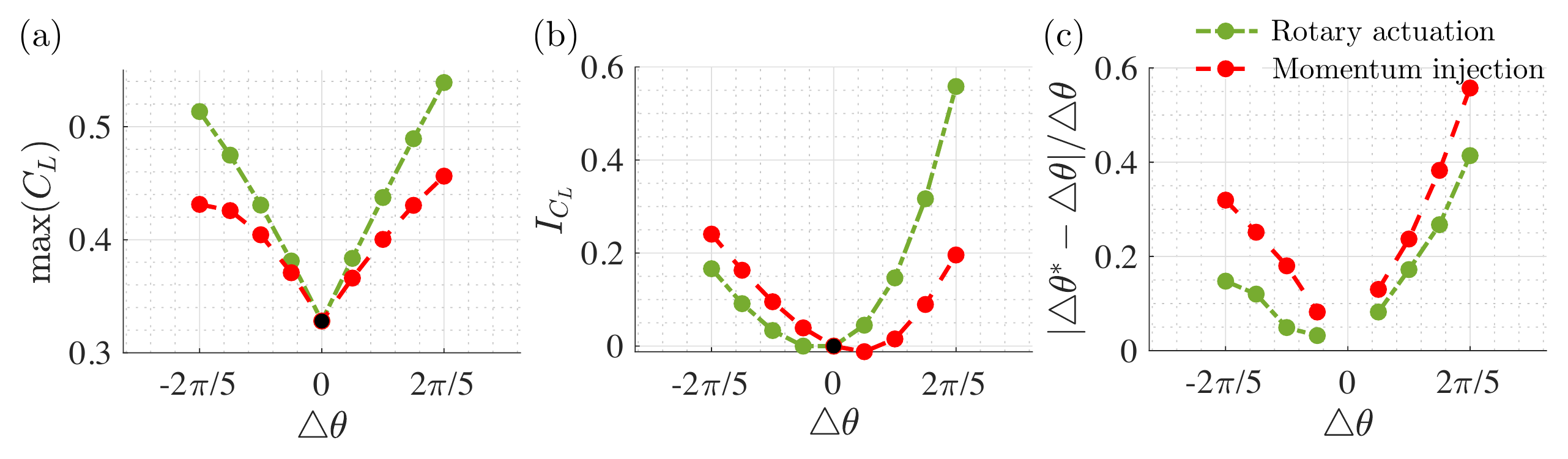}}
\vspace{-.125in}
\caption{Comparison of control performance for rotary actuation and momentum injection: (a) Maximum lift coefficient, (b) lift impulse and (c) error in asymptotic phase shift for varying desired phase shifts.}
\label{fig7}
\end{figure}

We now compare key features of interest for both rotary actuation and momentum injection. 
In particular, we investigate the maximum lift coefficient, lift impulse, and the error in asymptotic phase shift with phase optimal control, as shown in Figure~\ref{fig7}(a), (b) and (c), respectively. 
Lift impulse is the area under the lift-time curve~\cite{williams2009lift}, demonstrating the overall benefit of control in terms of lift enhancement.
With an increase in the phase shift, the maximum lift increases. 
With larger phase shifts, the benefit in lift enhancement increases dramatically. 
However, because of the considerable amplitude variation, the error in asymptotic phase shift increases with phase shift.
This is expected as the optimal control input is based on infinitesimal phase sensitivity functions and larger control amplitudes introduce nonlinear effects. 
Note that there is negligible error in the immediate transient phase shift.
The phase control performance of rotary control is superior to that of momentum injection from an aerodynamic view point. 
However, as mentioned before, the cost associated with rotary control is an order of magnitude higher.

\begin{figure}
\vspace{-.125in}
\centerline{\includegraphics[width=1.0\textwidth]{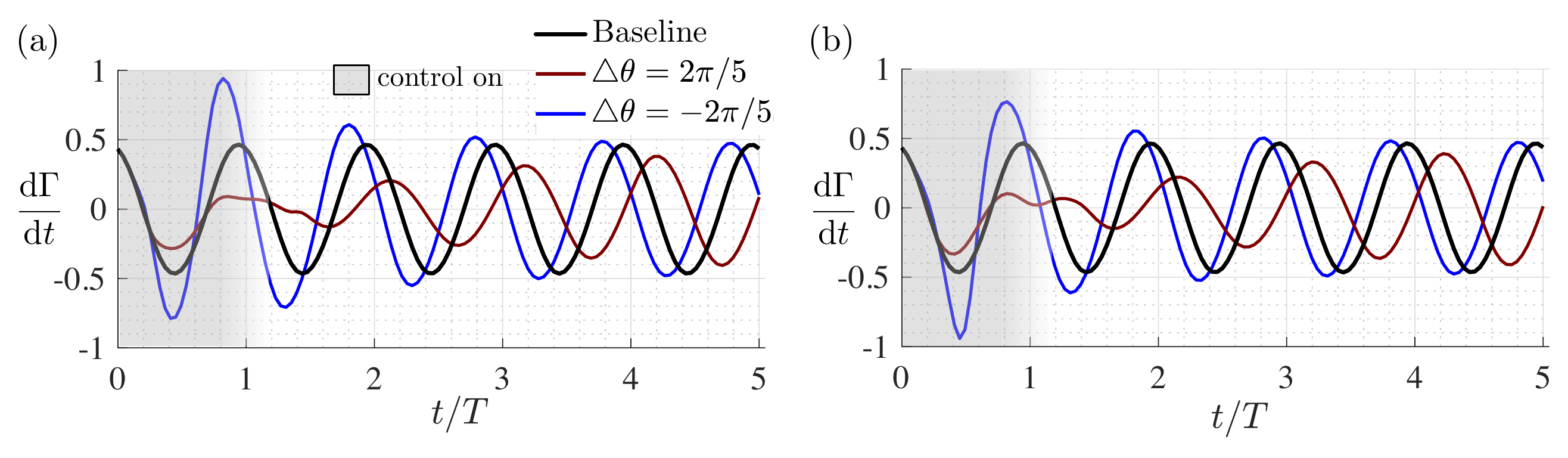}}
\vspace{-.125in}
\caption{Vorticity flux comparison with respect to the baseline for (a) rotary control and (b) momentum injection.}
\label{fig8}
\end{figure}

\citet{amitay2002controlled, amitay2006flow} analyzed transients associated with flow reattachment and separation using pulsed amplitude modulation of a synthetic jet actuator. 
They observed that shedding of large vortices and momentary increase in circulation were crucial in terms of achieving the transient flow attachment.
These enhancements were accentuated with actuation frequency close to the natural frequency of the flow~\cite{amitay2006flow, colonius2011control}. 
We similarly investigate flow transients in terms of the rate of change of circulation to characterize the effect of control given by
\begin{equation}
\frac{\mathrm{d}\Gamma}{\mathrm{d}t} = \int_{x/d = \text{const}} \omega(\boldsymbol{x},t) \boldsymbol{q}(\boldsymbol{x},t)\cdot \hat{\boldsymbol{e}}_x  \mathrm{d}y
\end{equation}
where $\boldsymbol{q}$ is the velocity field and $\omega$ is the vorticity. 
We compute the vorticity flux at $x/d = 1.2$.  
The rate of change of circulation in time, for both rotary actuation and momentum injection, are shown in Figure~\ref{fig8}(a) and (b)  for extreme phase shifts.
The trends are similar to the variation in the lift coefficient. 
The instantaneous vorticity flux for positive phase shifts is slightly greater than that of the baseline in the initial oscillation leading to an immediate increase in lift coefficient. 
The time-rate of change of circulation increases dramatically towards the end of the control forcing for the negative phase shift. 
This time corresponds to the vortex pinch-off, i.e., when the vortex is shed. 
The increase in flux is larger for the rotary actuation, leading to higher control performance compared to that of momentum injection.

\subsection{Flow over an airfoil}
\label{subsec:AF}

Finally, we demonstrate the optimal phase control approach on a two-dimensional, incompressible flow over a NACA0012 airfoil at an angle of attack $\alpha = 9^\circ$ with chord-based Reynolds number $Re = 1000$. 
This example is chosen to demonstrate the advanced capabilities of the phase-based control approach to enhance aerodynamic characteristics of wings in unsteady flow conditions.

\begin{figure}
  \centerline{\includegraphics[width=1.0\textwidth]{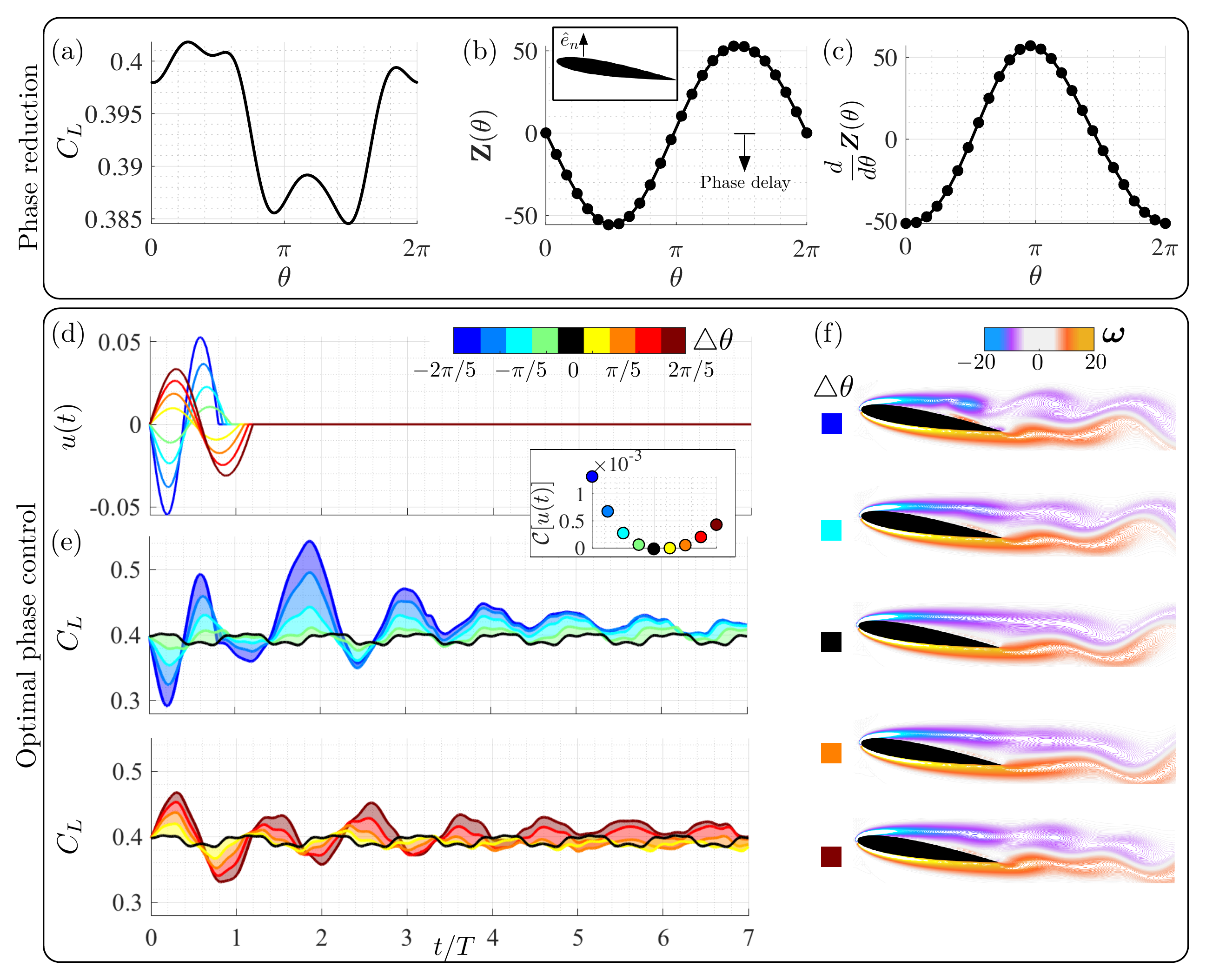}}
\vspace{-.1in}
\caption{Overview of phase-reduction and optimal phase control for flow over an airfoil using momentum injection: (a) Lift coefficient ($C_L$) of baseline cylinder flow with respect to the phase coordinate $\theta$,  (b) phase sensitivity function and (c) its gradient (right) for impulse perturbation introduced at an actuator located near the top separation point in the $\hat{e}_x$ direction, (d) energy-optimal control input needed to shift the phase of oscillation (with associated cost function in inset), (e) lift characteristics after application of forcing in direct numerical simulations for negative and positive phase shifts (shown separately), and (f) instantaneous flow fields after control is turned off ($t = 2T^*$). The color of the lines correspond to the cases with desired phase shift $\triangle \theta$ with $\triangle \theta = 0$ indicates baseline. For visual clarity, the fluctuation in the controlled lift trajectories are shown as shaded areas relative to the baseline lift coefficient.}
  \vspace{-.1in}
\label{fig9}
\end{figure}

The immersed-boundary projection method is used for direct numerical simulation~\citep{Colonius:CMAME08}. 
Five nested domains are used, with an outermost domain of $-16 \le x/c \le 16, -16 \le y/c \le 16$ and an inner grid resolution of $\triangle x = 0.0055$, where $c$ is the chord length. 
The integration time step is $\triangle t = 0.001$.
The baseline lift coefficient fluctuation is shown in Figure \ref{fig9}(a). 
The mean lift force is $\bar{C}_L = 0.3939$ and the oscillation frequency is $\omega_n = 2\pi f_nU/c = 7.17$, corresponding to a characteristic time of $T = 2\pi/\omega_n$. 
There are three fundamental differences in this high-dimensional flow example compared to the previous cylinder flow example: (i) the baseline mean lift force is nonzero, (ii) the harmonics of the natural shedding frequency are stronger and can be observed in the baseline lift oscillation, and (iii) the flow exhibits both leading-edge and trailing-edge separation.  

For phase-reduction and phase optimal control, we use wall-normal momentum injection, modeled by a localized force added near the average separation point.
We then perform the impulse-response analysis at $25$ characteristic phases $\theta$. 
The difference between the perturbed trajectories and the baseline fluctuations after $20$ time periods of oscillation is assessed to compute the phase sensitivity function, as shown in Figure~\ref{fig9}(b). 
After introducing the perturbation, the flow takes more oscillations to reach steady state compared to the cylinder flow. 
The amplitude of the perturbation is $\epsilon = 0.0005$.
The magnitude of the phase sensitivity function is larger compared to both actuation cases for the cylinder flow example.
The maximum phase advance sensitivity is observed at the phase corresponding to minimum lift coefficient. 
Phase delay sensitivity is largest at phases corresponding to larger lift coefficient.
Surprisingly, the phase sensitivity function has a sinusoidal shape, even for this highly complex flow. 
The corresponding gradient is shown in Figure~\ref{fig9}(c). 
The optimal control inputs for a desired phase shift $-2\pi/5 \le \triangle \theta \le 2\pi/5$ are shown in Figure \ref{fig9}(d). 
The associated cost of control (shown in inset) is lower than the corresponding cost of momentum injection for the cylinder flow example. 
This is expected as the magnitude of the phase sensitivity function for the airfoil flow is larger. 
The steady and unsteady coefficient of momentum for $\triangle \theta = -2\pi/5$ are $\bar{C}_\mu = 0.01$ and $C^\prime_\mu = 0.0132$, respectively. For $\triangle \theta = 2\pi/5$, $\bar{C}_\mu = 0.01$ and $C^\prime_\mu = 0.01$.

After applying the optimal forcing input to the Navier--Stokes equations and performing direct numerical simulations, we examine the fluctuations in the lift coefficient compared to the baseline for all cases in Figure~\ref{fig9}(e). 
To highlight the variation with respect to the baseline, we separate the lift fluctuation for negative phase shifts and positive phase shift. 
We also color the increments in the fluctuation to further emphasize the changes with control. 
Similar to the cylinder flow example, we achieve an immediate increase in lift for the positive phase shift cases. 
Immediately after control is switched off, we see a rapid increase in lift fluctuation for the negative phase shift cases. 
The transient phase shift is exactly as prescribed, indicating that immediate lock-on is obtained even for this complex flow. 
We also see that lift fluctuations are higher than the baseline for all cases for a number of oscillation cycles, even after the control is switched off. 
Note that although there are distinct harmonic oscillations in the baseline lift, when control is applied, these harmonics are suppressed until downstream.
The work of~\citet{joe2011feedback} on an inclined flat plate showed that larger ranges of negative phase shift in the control signal contributed to lift enhancement, while sharp decrease in enhancement at positive phase shift regimes were observed.
Although there is a preference for negative phase shifts for larger lift enhancement, the current phase control strategy enables lift enhancement over both positive and negative phase shifts.
While not shown, the drag force decreases during control for both positive and negative phase shifts. 
When the control is switched off, the drag force increases due to the nonlinear interaction with the trailing edge vortices, breaking the flow stabilization in the wake.

\begin{figure}
  \centerline{\includegraphics[width=0.65\textwidth]{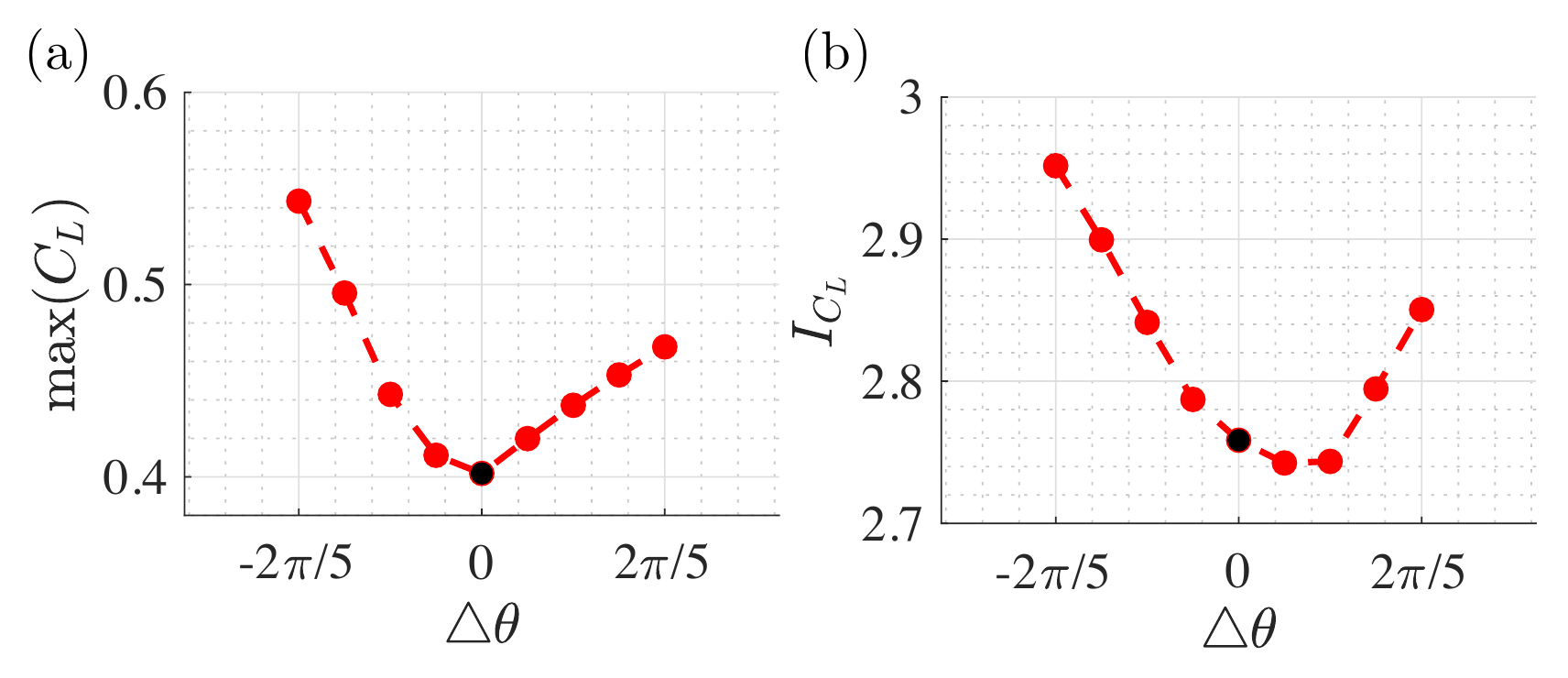}}
\vspace{-.1in}
\caption{Evaluation of control performance: (a) Maximum lift coefficient and (b) lift impulse for varying phase shifts.}
\label{fig10}
\vspace{-.1in}
\end{figure}

On observing the characteristic vorticity flow fields associated with the control cases in Figure~\ref{fig9}(f), we see that a leading-edge vortex (LEV)~\cite{wang2013low,Eldredge2019arfm} forms earlier for the controlled flow with negative phase shifts.
For positive phase shifts, the roll-up process of the leading edge vortex is delayed,  resulting in a wake elongation.
In the negative phase shift cases, the interaction of the shed LEV with the trailing edge vortex (TEV) results in the high lift after the control is switched off. 
These results are consistent with the findings of~\citet{darabi2004active} and~\citet{colonius2011control}.
Thus, with optimal phase control, we can precisely control the timing of vortex shedding. 
We also highlight the maximum lift force and corresponding lift impulse in Figure~\ref{fig10}(a) and (b), respectively. 
We can clearly see that negative phase shifts are preferred for maximum lift and lift impulse. 
A clear benefit in lift impulse for positive phase shifts are only observed for $\triangle \theta>\pi/10$. 
Thus, phase optimal control is an effective way of increasing transient aerodynamic performance. 

\section{Conclusions}
\label{sec:conclusion}

In this work, we developed a simple yet robust control strategy to modify the phase of highly nonlinear, periodic fluid flows.  
Our control strategy combines  phase-amplitude reduction and optimal phase-based control techniques for high-dimensional unsteady fluid flows. 
The phase-reduction technique used here provides a path to concisely describe the  phase dynamics of unsteady fluid flows via a phase sensitivity function, complementing Floquet theory.
Importantly, we show how this description can be used for effective flow control, demonstrating how it can be used to modify vortex dynamics in the near wake of bluff body flows, and on timescales that are commensurate with the period of vortex shedding. 
We consider several example systems, including the Stuart-Landau oscillator, the canonical fluid flow past a cylinder, and a highly complex fluid flow past an airfoil, illustrating the use of this approach for practical tailoring of aerodynamic responses. 
Our control strategy requires only measurements of physically attainable observables, such as lift and drag, and does not require high-dimensional data.

We demonstrate the performance of this phase-based control approach for two high-dimensional flow systems, the canonical flow over a cylinder and flow over an airfoil. 
In each system, our control law is only active for around one oscillation cycle, and is highly effective in modifying the phase of the flow to a desired phase shift.  
We explore the phase-optimal control strategy for a range of prescribed phase shifts with respective to the baseline oscillation. 
Interestingly, two separate mechanisms are exploited to achieve lift enhancement for negative and positive phase shifts. 
For negative phase shifts, the vortex formation process and shedding is accelerated; once the vortex is shed, we see a transient increase in lift. 
In contrast, for positive phase shifts,  vortex structures in the wake are elongated, causing an immediate increase in lift. 
As the wake is elongated and in turn stabilized, we also achieve a transient drag reduction.
Investigating the relationship between the control forcing and the phase-sensitivity function, we observe that for positive phase shifts, the control principle is similar to performing direct opposition control to restrict the shedding of vortex structures.

We propose two control approaches to rapidly and efficiently alter the phase dynamics of unsteady fluid flows, and comparing these two approaches illuminates some interesting trade-offs.  
In the first approach, we solve an offline optimization problem using the Euler-Lagrange equations and then design the optimal forcing. 
Since this solution is tied directly to the phase sensitivity function, it is simple, effective, and has an intuitive interpretation.
The second approach involves real-time feedback control using model-predictive control (MPC). 
For the simple Stuart-Landau oscillator, the two approaches are remarkably similar.
However, we show that the MPC approach is especially successful in systems with stochastic disturbances and slowly varying parameters, where MPC produces effective control laws that correct for model uncertainty and are robust to noise.

This work is mainly focused on controlling the phase of periodic vortex shedding in unsteady fluid flows, and there is tremendous scope for future work to build on this analysis. 
Coupling the current framework with isostable coordinates~\cite{wilson2020data}, it will be possible to alter both phase and amplitude simultaneously for unsteady fluid flows.
Extending this work for quasi-periodic and turbulent flows is an important future direction.  
The phase-amplitude reduction analysis could be combined with cluster-based approaches~\cite{kaiser2013cluster, nair2019cluster} to improve its applicability to  aperiodic and turbulent flows.  
Further, demonstrating this approach in three-dimensional experimental configurations will be critical to its wide adoption.  
Finally, it may be possible to extend the optimization procedure described above to a longer horizon, potentially uncovering more favorable control laws.

\section*{Acknowledgements} SLB acknowledges support from the Army Research Office (W911NF-17-1-0306) and the Air Force Office of Scientific Research (FA9550-18-1-0200). BWB and SLB acknowledge support from the Air Force Office of Scientific Research (FA9550-19-1-0386). BWB acknowledges support from the Air Force Office of Scientific Research (FA9550-18-1-0114) and the Air Force Research Lab (FA8651-16-1-0003). 
KT acknowledges support from the Air Force office of Scientific Research (FA9550-18-1-0040). We would also like to thank Prof. Hiroya Nakao, Dr. Chi-An Yeh, and Bharat Monga for valuable discussions. 

\newpage
\section*{Appendix: Model-predictive control (MPC)}
\label{subsub:MPC}

\begin{figure}
  \vspace{-.1in}
  \centerline{\includegraphics[width=.950\textwidth]{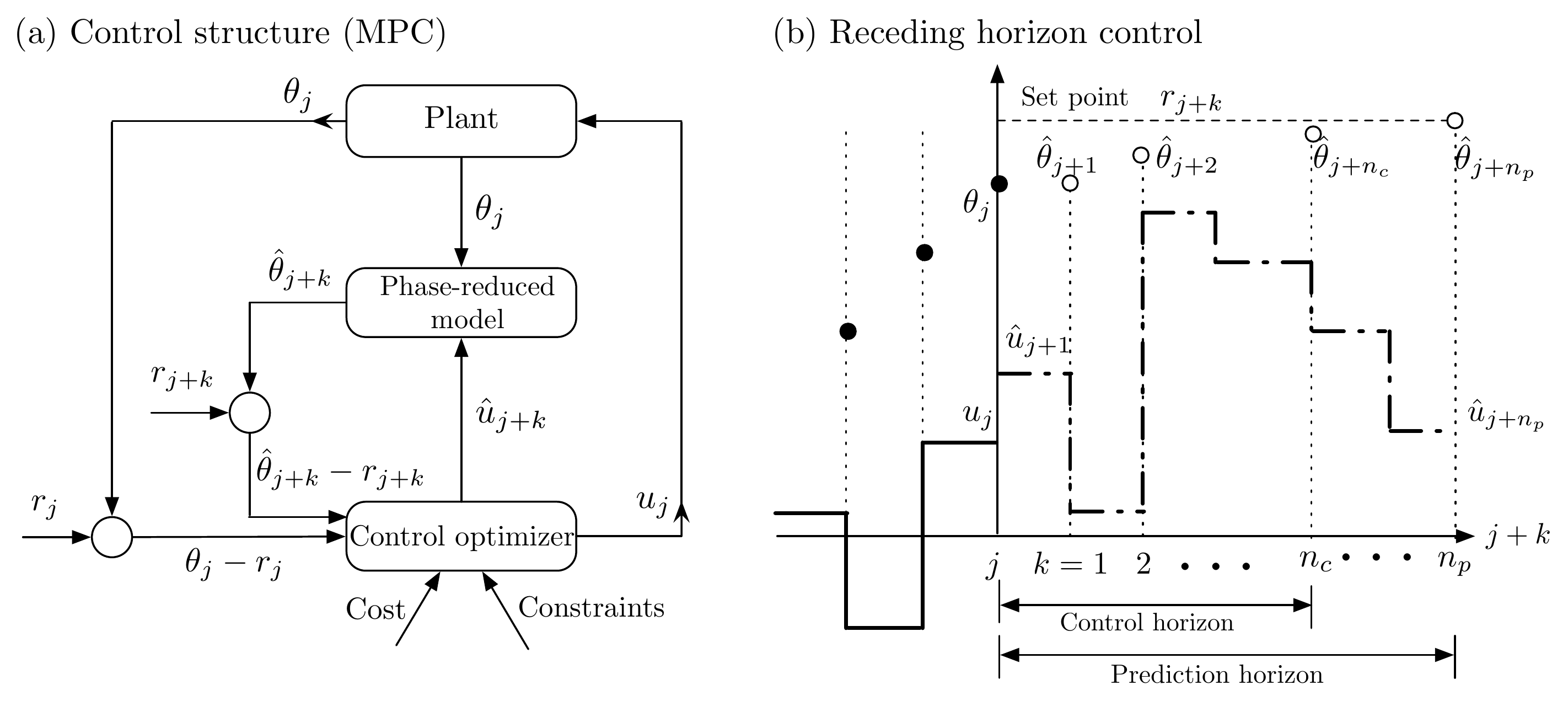}}
  \vspace{-.1in}
  \caption{(a) Schematic of model-predictive control (MPC) structure and (b) receding horizon framework (adapted from~\citet{kaiser2018sparse}).}
\label{fig11}
\end{figure}

Here, we elaborate on model-predictive control (MPC)~\cite{garcia1989model} for optimal phase control.  
MPC can be used to perform real-time feedback control, can incorporate constraints, and has been shown to be robust to noisy observations. 
A schematic for MPC is shown in Figure~\ref{fig11} (a). 
For a problem of interest, there are three key components for MPC: (i) the dynamical system (plant) on which control is to be performed (in the case of fluids flows, this would be the Navier--Stokes equation, given by Eq.~(\ref{eq1})), (ii) the model used for prediction of the state (or observable), and (iii) the control optimization algorithm used to deduce the real-time optimal control law. 
We also require knowledge of a desired reference state to be achieved with control. 
In the MPC framework, an optimal control problem is solved over a receding horizon to determine the next control input. 
The optimization is repeated at each time step and the control law is updated, as shown in Figure \ref{fig11} (b). 

In the work of~\citet{kaiser2018sparse}, MPC was combined with several models based on dynamic mode decomposition (DMD)~\cite{Schmid2010jfm,proctor2016dynamic}, neural networks~\cite{hornik1989multilayer} and sparse identification of nonlinear dynamics (SINDy)~\cite{brunton2016discovering}. 
Here, we use the phase-reduced model, given by Eq.~(\ref{eq4}), where $\boldsymbol{f}(t)$ is the forcing input. 
For simplicity, at time $t_j$, $r_j$ is the reference state, $\theta_j$ is the phase coordinate, and $u_j$ is the control input. 
Note that for unsteady fluid flows, the forcing is described by Eq.~(\ref{eq9}). 
For phase-based control, the reference $r$ should be $T^*$ periodic for a desired phase shift of $\triangle \theta = \omega_n(T^* - T)$; thus $r_j = 2\pi t_j/T^*$. 
The output of the phase-reduced model is $\hat{\theta}$ and the control optimizer output is $\hat{u}$. 
As shown in Figure \ref{fig11} (b), the prediction and optimization horizons are $n_p$ and $n_c$ time steps, respectively.
At each step, the control optimization identifies a sequence of control inputs $\hat{u} = \{\hat{u}_{j+1}, \hat{u}_{j+2}, \cdots ,\hat{u}_{j+n_c}\}$ based on predicted phases $\hat{\theta} = \{\hat{\theta}_{j+1}, \hat{\theta}_{j+2}, \cdots ,\hat{\theta}_{j+n_p}\}$, after which $u_j = \hat{u}_{j+1}$ is applied to the plant, and then the procedure is repeated.

The MPC optimization~\citep{kaiser2018sparse} at each time step is given by
\begin{equation}
\mathop{\text{min}}_{\substack{\hat{u}}} \mathcal{J}= \mathop{\text{min}}_{\substack{\hat{u}}} \sum_{\substack{k=0}}^{n_p} ||\hat{\theta}_{j+k} - r_{j+k}||^2_{Q} +  \sum_{\substack{k=1}}^{n_c} (||\hat{u}_{j+k}||^2_{R_u} + ||\triangle \hat{u}_{j+k}||^2_{R} )
\label{eq6}
\end{equation}
subject to constraints $\hat{u}_\text{min} \le \hat{u}_k \le \hat{u}_\text{max}$ and $\triangle \hat{u}_\text{min} \le \triangle \hat{u}_k \le \triangle \hat{u}_\text{max}$. 
Here $\triangle \hat{u}_k = \hat{u}_k - \hat{u}_{k-1}$ is the input rate. 
We penalize the phase coordinate, control input and its rate. 
Also, for each term, a weighted norm is computed $||x||_Q = x^T Q x$. 
The weights are chosen to be positive semi-definite.

\bibliographystyle{jfm}
 \begin{spacing}{.8}
 \small{
 \setlength{\bibsep}{3.pt}
 \bibliography{phase_control_references}
 }
 \end{spacing}

\end{document}